\documentclass[sigconf,10pt]{acmart}

\usepackage[skip=-2pt]{subcaption}
\usepackage{tikz}
\usepackage{url}

\usepackage{breakurl}
\usepackage{caption}
\usepackage{color}
\usepackage{xspace}
\usepackage{comment}
\usepackage[export]{adjustbox}
\usepackage[english]{babel}
\usepackage{enumitem}
\usepackage{diagbox}
\usepackage{slashbox}
\usepackage{array}
\usepackage{multirow}
\usepackage{wasysym}
\usepackage{xcolor,colortbl}
\usepackage{cleveref}
\usepackage[super]{nth}
\usepackage{makecell}
\usepackage{hyperref}
\hypersetup{
    colorlinks=true,
    linkcolor=red,
    filecolor=red,      
    urlcolor=red,
    citecolor=red
    }

\setlist{itemsep=0pt,parsep=0pt}

\newcommand{\totalHandovers}{23630\xspace}
\newcommand{\hsrHandovers}{17085\xspace}
\newcommand{\incpHandovers}{3947\xspace}
\newcommand{\miscHandovers}{2598\xspace}

\newcommand{\noPredictionAvaragePercentage}{1.7\%\xspace}

\newcommand{\highSpeedTrainAccuracy}{97.7\%\xspace}
\newcommand{\drivingTestAccuracy}{92.9\%\xspace}
\newcommand{\miscAccuracy}{92.4\%\xspace}
\newcommand{\averageAccuracy}{97\%\xspace}
\newcommand{\averageFalsePredictioins}{1.3\%\xspace}

\newcommand{\highSpeedTrainMaxAccuracy}{99.3\%\xspace}
\newcommand{\drivingTestMaxAccuracy}{99.5\%\xspace}
\newcommand{\miscMaxAccuracy}{99\%\xspace}

\newcommand{\sysname}{{EdgeWarp}\xspace}
\newcommand{\Name}{\textit{{\sysname}}\xspace}

\newcommand{\Old}{\textit{{Existing 5G}}\xspace}

\newcommand{\handover}{\textit{handover}\xspace}

\newcommand{\lowPriorityIot}{\textit{\Name-LP}\xspace}

\newcommand{\mediumPriorityIot}{\textit{\Name-MP}\xspace}

\newcommand{\highPriorityIot}{\textit{\Name-HP}\xspace}

\newcommand{\defaultPriority}{\textit{Existing 5G}\xspace}

\newcommand{\mobilityHintWindow}{100\,ms\xspace}

\newcommand{\redis}{{Redis}\xspace}
\newcommand{\memcache}{{Memcached}\xspace}

\newcommand{\uniformLowOverDefaultMedium}{\textsf{$2.1\times$}\xspace}
\newcommand{\burstyLowOverDefaultMedium}{\textsf{$2\times$}\xspace}

\newcommand{\repourl}{https://anonymous.4open.science/r/Mobicom25-EdgeWarp-3C83}

\renewcommand\footnotetextcopyrightpermission[1]{} 
\setcopyright{none}

\begin{document}



\author{Mukhtiar Ahmad, Faaiq Bilal, Mutahar Ali, Muhammad Ali Nawazish, Amir Salman, Shazer Ali, Fawad Ahmad, Zafar Ayyub Qazi}

\title{Warping the Edge: Where Instant Mobility in 5G Meets Stateful Applications}


\begin{abstract}
Edge computing is considered a key paradigm for supporting real-time applications over 5G networks, as hosting applications at the network edge can substantially reduce delays. A significant fraction of real-time applications over 5G are expected to be highly mobile applications. However, one challenge with hosting mobile applications on the network edge is ensuring that users continue to get low latency as they move across different locations. This requires the support to handover clients to different edge sites with negligible application delays. However, many edge applications are stateful and can experience significant downtime during state migration over 5G. This paper addresses the problem of enabling stateful mobile edge applications in 5G networks. We first identify the key architectural issues and then propose a new system design, \Name, that mitigates delays during mobility through proactive application state migration. To enable this, we extend the existing edge data stores with the design of a novel two-step application state synchronization protocol, that leverages the early prediction of the target edge host. Additionally, \Name prioritizes the handover of latency-sensitive edge applications by communicating their latency requirements to the 5G control plane at the beginning of a data session. Our evaluation with real edge applications shows up to a 15.4$\times$ reduction in application downtime under mobility. We have made our anonymized code publicly accessible~\href{\repourl}{\underline{here}}.
\end{abstract}

\maketitle
\pagestyle{plain}
\section{Introduction}
5G and beyond networks aim to support intriguing emerging latency-sensitive applications (apps) that promise to enhance safety and user experience. These apps encompass a range of use cases, such as connected and autonomous vehicles~\cite{ahmad_carmap, zhang2021emp, infra_structure_assisted_car_map}, virtual and augmented reality (AR/VR) \cite{elbamby2018toward, ar_vr_satyanarayanan2017emergence}, edge-based video analytics~\cite{killer_app_edge_computing}, and multi-player online gaming \cite{online_gamming_delay_budget,zhang2017towards_online_game_updates_rate}. Such apps have stringent delay budgets, e.g., VR apps with head tracking systems require latency of less than 17\,ms to achieve perceptual stability~\cite{elbamby2018toward, ar_vr_satyanarayanan2017emergence}. To enable support for such apps, 5G networks are integrating Edge Computing~\cite{ar_vr_satyanarayanan2017emergence,mec_in_5g}; a paradigm whereby apps are hosted closer to the users, on cell towers, cell aggregation sites, or other edge sites~\cite{mec_in_5g, etsi2, etsi3}. Many cellular providers are partnering with cloud providers for deploying edge apps inside the cellular infrastructure~\cite{edge_sites_american_tower,azure_edgezone, aws_wavelength, google_carriers}.  

Hosting apps at the network edge can significantly reduce delays. However, it also makes it challenging to ensure that users always receive low latency as they move across different locations. This, in turn, requires support to handover clients to different edge sites with negligible delays~\cite{mec_app_mobility_api_021}.
These app handovers are likely to be frequent~\cite{frequent_handover_zhou20215g, frequent_migration_abouaomar2021deep,  driving_frequent_app_handoverr,mec_in_5g, mec_evolution_toward_5g_etsi, etsi2} for a significant fraction of 5G use cases as more than 50\% of 5G IoTs are expected to be highly mobile apps related to the automotive industry \cite{5g_edge_computing_aws_53p}.\footnote{5G networks will have 5 billion connected IoTs by 2025~\cite{iot_in_5g_era_cisco}.} 
Many of these edge apps are stateful~\cite{zhang2021emp, ahmad_carmap, zhang2017towards_online_game_updates_rate, shi2019mobile_edge_vr, full_body_tracking_vr} and migrating stateful apps over 5G can cause significant app downtime during mobility~\cite{vm_container_migration_in_5g_mec,live_migration_yang2021multitier} (\S\ref{sec:slow_app_handover}). Consequently, if the app downtime exceeds the delay budget for latency-sensitive apps, it can compromise user safety in the case of connected cars~\cite{zhang2021emp, ahmad_carmap} or cause users motion sickness when using immersive mobile VR~\cite{motion_sickness_kundu2021study}.

In this paper, we first step back and ask the question, \textit{what are the key architectural bottlenecks in enabling stateful mobile edge apps over 5G}? We identify three key issues in enabling fast app handovers during mobility (illustrated in Figure~\ref{fig:challnges_serialized_ho}):

\begin{figure*}
\centering
\includegraphics[width=\textwidth]{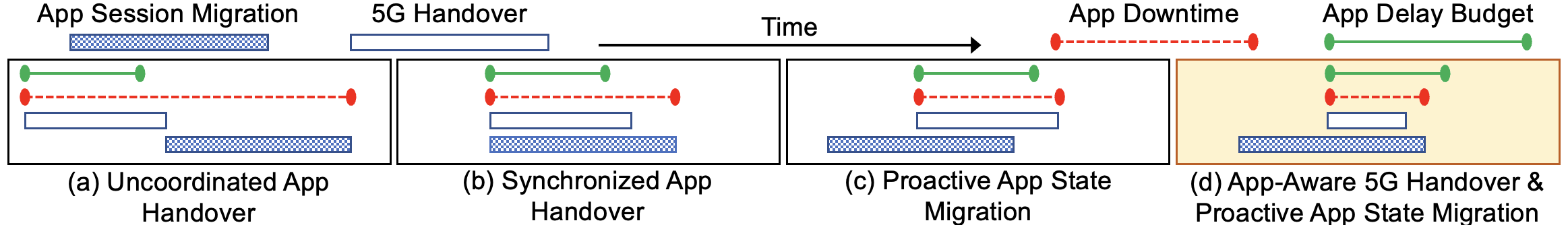}
\vspace{-0.3in}
\caption{Key challenges and potential solutions to minimize downtime during application handover caused by mobility. Typical 5G handover delays are in the order of 100 to 200\,ms~\cite{understanding_operational_5g}, while application session migration delays exceed 140\,ms with existing methods (\S\ref{sec:slow_state_migration}).}
\label{fig:challnges_serialized_ho}
\vspace{-0.2in}
\end{figure*}

\begin{itemize}[leftmargin=*]
\item \textbf{Uncoordinated and Reactive App Session Migration:} Stateful app handover involves two key steps: (i) the user's network session migration, also known as 5G handover\footnote{A 5G handover can also trigger user plane session migration---referred to as Packet Data Unit session migration~\cite{3gpp_123_502,pdu_session_establishment}).} and (ii) the user's app session migration. Uncoordinated migration of the network and app sessions can cause extended app downtime (Figure \ref{fig:challnges_serialized_ho}a). If the app session migration occurs after a 5G handover, it can result in two interruptions, one during the 5G handover and another during the app session migration. Even if app session migration is performed proactively, latency-sensitive apps that frequently update app state~\cite{zhang2017towards_online_game_updates_rate} can incur non-zero downtime to retrieve the latest app state. 
On the other hand, even if we perfectly synchronize 5G handover and app session migration events~(Figure~\ref{fig:challnges_serialized_ho}b), app downtime can still exceed the delay budget~(<~17\,ms~\cite{elbamby2018toward, ar_vr_satyanarayanan2017emergence}) of latency-sensitive edge apps.

\item \textbf{Slow App Session Migration:} A key challenge with speeding up session migration of stateful apps is that they typically require the user’s up-to-date app state at the target edge host before a user connects to it~\cite{harchol2018cessna}. However, existing techniques for app session migration are either slow~\cite{live_migration_8000803, vm_container_migration_in_5g_mec, hardware_accelerated_live_migration, efficient_container_migration, live_migration_269357, criu} or rely on app-specific optimizations~\cite{fast_check_point_recovery}~(\S\ref{sec:slow_state_migration}). 
Moreover, simpler session migration techniques, such as continuing the app session with the source instance before state migration completes, can result in two interruptions: one during the 5G handover and another during app session migration. 
This occurs because the app session at the source edge host needs to be frozen, migrated, and restarted on the target edge host to ensure state consistency.
The key challenge is to design a session migration technique that is \textit{fast} and \textit{generic}.

\item \textbf{App-agnostic 5G Handover:}
We observe that even if we somehow address the issue of slow app session migration, the time required for a 5G handover can become a bottleneck in enabling latency-sensitive edge apps, as shown in Figure~\ref{fig:challnges_serialized_ho}c (\S\ref{sec:app_agnostic_5g_handover}). 
An architectural issue is that 5G control plane processing is \textit{application-agnostic} and there are no 5G interfaces through which edge apps can inform the cellular control plane of their latency requirements.

\end{itemize}

To address the above issues, we propose a new system design, \Name, for edge apps over 5G. \Name is a cross-layer design that consists of the following key ideas: 

\begin{itemize}[leftmargin=*]
    
\item \textbf{Target BS Prediction:}
A coordinated and proactive approach to app handovers is essential to minimize app downtimes. To achieve this, \Name predicts the target BS\footnote{Predicting the target BS is equivalent to predicting the target edge host as it is a function of target BS. Details in \S\ref{sec:goals_and_assumptions}.} for a 5G handover in advance, allowing edge apps to initiate app state migration proactively (Figure \ref{fig:challnges_serialized_ho}c). 
Existing techniques for target BS prediction
rely on handover history~\cite{bs_prediction_ho_history, bs_prediction_low_accuracy_3}, LTE events~\cite{bs_prediction_low_accuracy_1}, or machine learning approaches~\cite{bs_prediction_low_accuracy_2, bs_prediction_low_accuracy_4, bs_prediction_radio_link_characteristics}. Among these, the 5G network-based approaches have two common limitations: (i) they lack early prediction timing, which is crucial for practical benefits, as predicting the target BS only a few milliseconds before a handover provides negligible performance gains (\S\ref{sec:eval:impact_of_mobility_hint}), and (ii) their prediction accuracy is generally poor (< 85\%).

To overcome these limitations, we propose a multi-step early target BS prediction pipeline (\S\ref{sec:design:target_bs_predction}), which leverages Radio Network Information (RNI) APIs in MEC~\cite{5g_mec_rni1} to predict handovers in advance using an LSTM network, then extrapolate radio measurements to enable early target BS prediction, and finally perform target BS prediction. We are using LSTMs because they excel at capturing both short-term fluctuations and long-term patterns in the rapidly changing time series data from 5G mmWave radio access networks~\cite{akdeniz2014millimeter}. We evaluate this pipeline on real 5G/4G radio network traces, including data from driving tests~\cite{incp_dataset}, high-speed trains~\cite{li2020beyond}, and various other mobility scenarios~\cite{mobile_insight_logs}. Our evaluation shows on average \averageAccuracy correct 
 predictions of the target BS 100 ms ahead of the handover~(\S\ref{sec:eval:target_BS_prediction}), which we refer to as a \textit{mobility hint} for apps.
Our approach outperforms existing techniques, including LSTM-based networks~\cite{bs_prediction_low_accuracy_4}, by leveraging (i) enhanced feature vectors 
capturing time-series radio signal trends 
extrapolation of radio signals (\S\ref{sec:design:target_bs_predction}) when sample data is insufficient.
 

\item \textbf{Two-Step App State Synchronization:}
An app can proactively synchronize all or a subset of the user's app state with the target edge host based on mobility hint. However, synchronizing a highly dynamic state object that is frequently updated offers little advantage as it quickly becomes stale. Additionally, identifying and tracking different types of state objects for proactive synchronization is a daunting task for developers. Instead, we propose extending existing edge data stores (e.g., \redis~\cite{redis_store} and \memcache~\cite{memcached_store}) to store and track the app state of an edge app. This extension allows us to track the update rate for each state object of an edge app~(\S\ref{sec:design_app_state_migration}). The server side of edge apps stores the per-user (in-memory) app state in our modified external data store. The modified data store exposes new APIs 
~(Table~\ref{tab:api_in_edge_store}) to edge apps to enable the synchronization of state objects with low update rates upon receiving a mobility hint, and the remaining app state is synchronized during a 5G handover to maximally overlap interruption time (Figure~\ref{fig:challnges_serialized_ho}c). This approach reduces blocking time during state migration, resulting in minimized app downtime~(\S\ref{sec:state_migration_delay}).

\item \textbf{App-Aware Control Plane Processing:}
To make 5G control plane processing app-aware, we introduce new APIs (Table \ref{tab:api_for_app_influence_on_cpf}) that allow edge apps to report their latency requirements and prioritize control plane processing accordingly (\S\ref{sec:app_impact_on_control_plane}). These APIs are called by the server side of latency-sensitive apps through the MEC \cite{mec_in_5g}, which influences control plane processing. Restricting API access to the server side ensures better control and prevents misuse, as it is under the network operator's control or managed by trusted third parties. Our experiments demonstrate that an app-aware 5G control plane significantly reduces handover delays, bringing edge app downtime within their specified delay budgets (\S\ref{sec:app_aware_control_plane_processing}).

\end{itemize}

To show the efficacy of the above ideas, we implement \Name. We implement a multi-step target BS prediction pipeline utilizing an 8-layer stacked LSTM neural network on top of Keras library~\cite{keras_library}. Our data sets consist of real 4G/5G traces with a total of \totalHandovers handovers. On the app layer, we implement dynamic state tracking and two-step state synchronization in \redis~\cite{redis_store} and \memcache~\cite{memcached_store} data stores~(\S\ref{sec:design_app_state_migration}). To evaluate the impact of \Name on real edge apps, we modify two recent open-source edge apps, CarMap~\cite{ahmad_carmap} and EMP~\cite{zhang2021emp}, to use our modified \redis data store for proactive state synchronization. As real apps provide limited options to vary app state properties (e.g., state sizes, update rate, etc.), we implement a simulation framework in Python to simulate the behavior of a wide variety of latency-sensitive edge apps~(\S\ref{sec:imp_modified_state_store}). We interface a Python version of both modified \memcache and \redis key-value data stores with the simulation framework to evaluate the impact of two-step state syncing on a wide variety of app behaviors. To implement app-aware control plane processing, we modify a recently proposed cellular control plane~\cite{neutrino_ton}. We provide further details of our implementations in \S\ref{sec:implementation}. We open-source our code on an anonymous Github link: \href{\repourl}{\repourl}.

Our evaluation shows up to $14\times$ and $15.4\times$ improvement in app downtime for CarMap and EMP, respectively.
We correctly predict the target BS in \averageAccuracy cases at least 100\,ms in advance, in \averageFalsePredictioins cases less than 100\,m       s in advance, and fail to predict the correct target BS in \noPredictionAvaragePercentage cases. Apps can significantly improve performance during mobility with correct target BS predictions (Figure \S\ref{fig:challnges_serialized_ho}c), while neither improving nor degrading performance in the case of false predictions due to fallback on reactive migration (Figure \S\ref{fig:challnges_serialized_ho}b).
Consequently, both CarMap and EMP show more than $2.5\times$  improvement in app state migration time~(\S\ref{sec:state_migration_delay}). 
\Name shows up to $10^3\times$ improvement in median 5G handover completion time for latency-sensitive apps~(\S\ref{sec:app_aware_control_plane_processing}).  

\noindent\textbf{Contributions:} In summary, we make the following contributions to this work: 
\begin{enumerate}[leftmargin=*]
\item We identify the key architectural bottlenecks in enabling stateful mobile apps over 5G (\S\ref{challenges}).
\item We design a new system architecture, \Name, compatible with existing RAN and cellular core (e.g., 5G/4G/LTE). \Name consists of three new modules to speed up app handovers (\S\ref{sec:design}).
\item We implement \Name. Our implementation consists of a target BS predictor, an extended edge data store on top of Redis and Memcached, and an app-aware 5G control plane (\S\ref{sec:implementation}).
\item We evaluate \Name with two real edge apps, CarMap, and EMP, with multiple real-world cellular data sets, and also perform a sensitivity analysis using a simulation framework (\S\ref{sec:evaluation}).
\end{enumerate}

\section{Background and Motivation}

\subsection{5G Architecture}
\label{sec:5g_background}

\begin{figure}
\centering
\includegraphics[width=0.65\columnwidth]{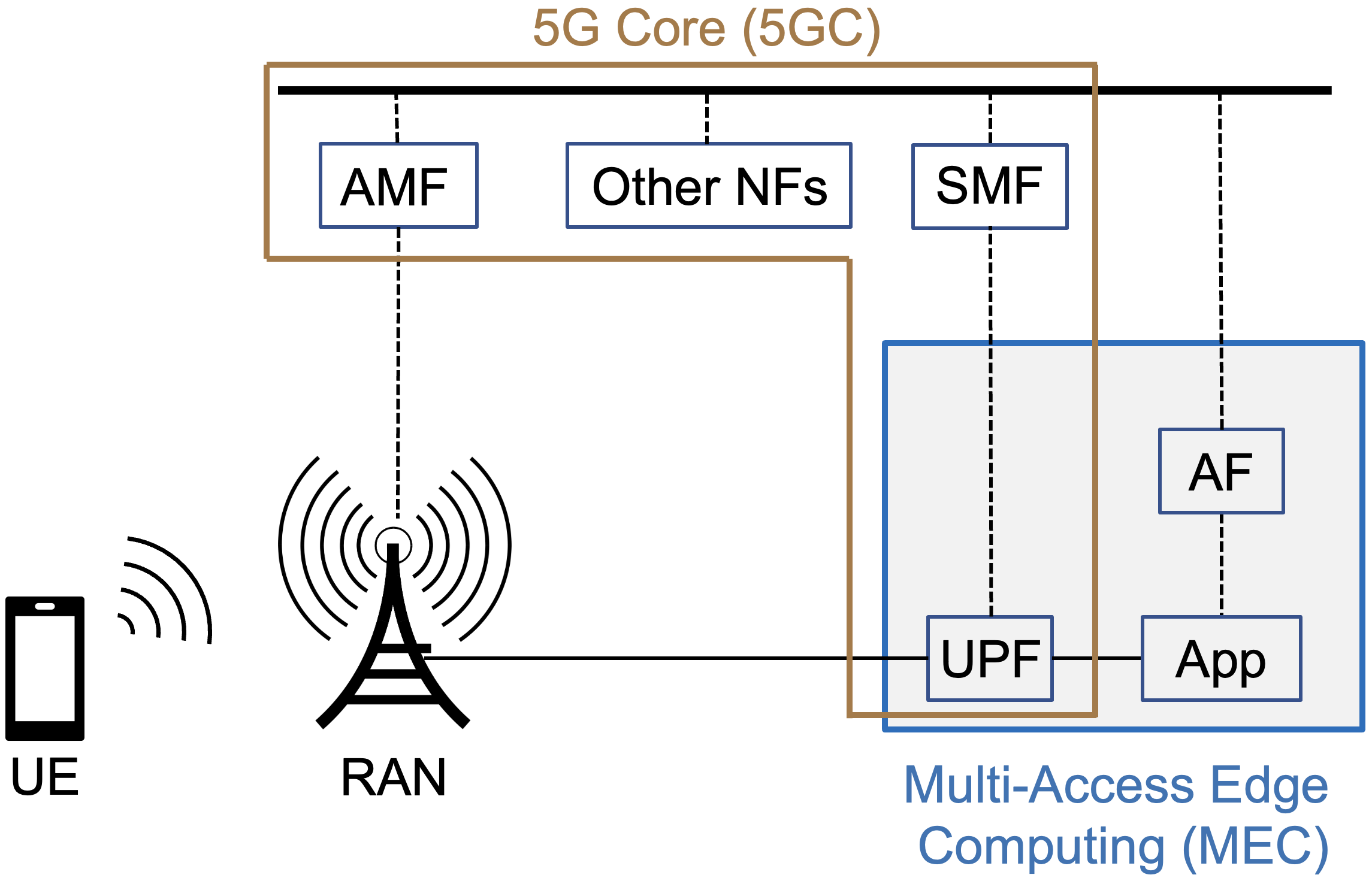}
\vspace{-0.1in}
\caption{Simplified 5G network architecture.}
\vspace{-0.22in}
\label{fig:5g_simplified_network_architecture}
\end{figure}

A 5G network consists of User Equipment (UE), Radio Access Network (RAN), and a 5G Core (5GC), as shown in Figure~\ref{fig:5g_simplified_network_architecture}. 5GC is decomposed into virtualized microservices in the service-based architecture introduced by 5G.
The service-based 5G architecture also enables the concept of network slices \cite{applied_network_slices_in_5g, ericsson_network_slice_for_gaming, ericsson_e2e_network_slices}. A 5G network can be configured to produce different logical networks (network slices) on top of shared physical infrastructure for different classes of app traffic. Below we describe the key 5G core functions.

\noindent\textbf{Data Plane Functions:}
The User Plane Function (UPF) performs the key data plane functions such as connecting base stations to external data networks (DN), IP address allocation, traffic routing, usage reporting, and buffering of downlink traffic. One or more UPFs can serve a single user.

\noindent\textbf{Mobility and Session Management Functions:}
The Access and Mobility Management Function (AMF) performs the key access and mobility-related functions, including access authentication, access authorization, and mobility management. Every UE is served by a  single AMF. The Session Management Function (SMF) performs session creation, modification and release, UPF selection, and downlink data notification. One or more SMFs can serve a single UE.

\subsection{Edge Computing Support in 5G} 
\noindent\textbf{Multi-access Edge Computing (MEC):} 
MEC is a system architecture aimed to provide storage and processing to  mobile users at the network edge for low latency and bandwidth efficiency \cite{mec_in_5g}. The 5G system architecture \cite{3gpp_architecture} allows the MEC system to collaboratively interact with the 5G system in traffic routing and policy control. For instance, the MEC system can influence the selection or re-selection of an appropriate UPF to route data traffic to a suitable edge app. The 5G network can also expose real-time radio network information to apps through the MEC.

\noindent\textbf{Application Function:} 
MEC allows apps through the Application Function (AF) \cite{3gpp_architecture} to route data traffic to the targeted apps at the edge sites. Since a UPF plays a central role in traffic routing, an edge app through an AF can influence the 5G core network to select a particular UPF for the UE~\cite{3gpp_architecture}. An AF can also influence 5G core network policies and expose services to the users. 

\subsection{Stateful Edge Applications}
\label{sec:stateful_edge_apps}
5G networks aim to support a large number of edge apps with MEC~\cite{etsi2}. These apps can be either stateless or stateful. By stateful apps we mean apps whose processing depends on the state previously reported by the client and/or generated at the server.
An example of a simple stateful app is an edge server maintaining a count of the number of packets received from a client~\cite{harchol2018cessna}.
On user mobility, the user app state (e.g., packets count) will be required at the new edge server (to maintain a correct packets count) before a user connects to it. Real-life stateful edge apps can be much more complex. Below, we discuss a few of these stateful edge apps that are latency-sensitive and can be mobile.

\noindent\textbf{Real-Time Mapping and Localization:}
Aerial and ground robots~(e.g., CarMap~\cite{carmap_gitub}) build real-time 3D maps of the environment and use these maps to accurately position themselves with respect to their surroundings~\cite{edge_slam, slam_share,swam_map,carmap_gitub}. These 3D maps are updated using onboard depth perception sensors, such as stereo cameras and LiDARs. These maps are typically maintained at the network edge for real-time localization in a dynamic environment~\cite{xu2022swarmmap_edge_car_map}.

\noindent\textbf{Autonomous Vehicles:} 
Autonomous vehicles can leverage edge computing and sensing capabilities, such as infrastructure-mounted LiDARs and stereo cameras, for safer and more efficient driving~\cite{wang2018networking,biswas2023autonomous,zhang2021emp}.
EMP~\cite{zhang2021emp} is an  example of edge-assisted multi-vehicle
perception system for connected and autonomous vehicles (CAVs). Such systems merge the raw sensor data shared by multiple CAVs on an edge server to enhance the perception of participating CAVs to improve road safety and traffic efficiency.

\noindent\textbf{Mobile VR:}
Apps such as those based on augmented reality (AR) and virtual reality (VR) require ultra-low latency in the order of a few milliseconds \cite{vr-latency} to provide perceptual stability. Mobile devices running VR Apps can offload computation to MEC for efficiency \cite{etsi2}. Edge-based VR services \cite{cloud_vr_network, shi2019mobile_edge_vr, full_body_tracking_vr} are examples of such Apps. 

\noindent\textbf{Online Gaming:}
Massive multiplayer online gaming is another example of stateful services at the network edge to achieve ultra-low-latency \cite{zhang2017towards_online_game_updates_rate, online_gaming_1, online_gaming_2}. Such apps can have highly dynamic app state (i.e., > 100 updates/s) and latency budget < 20\,ms.  

\subsection{Slow Application Handover}
\label{sec:slow_app_handover}

App handover for latency-sensitive edge apps should be performed with minimal delay to reduce app downtime.\footnote{The duration for which an app is unable to serve the user.} 
The time taken for an app handover depends on many factors, including user-specific app state size, state update frequency, app code size, and available network bandwidth. 

\begin{figure}[]
\centering
\begin{minipage}[t]{0.48\linewidth}
\includegraphics[width=\columnwidth]{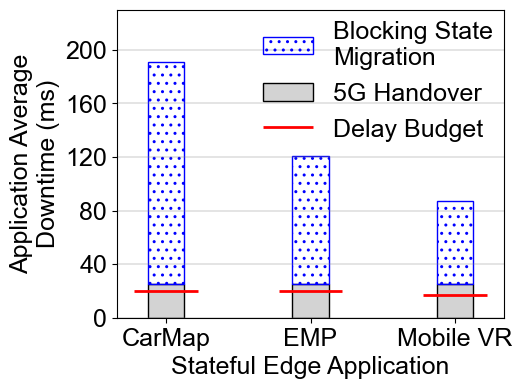}
\vspace{-0.3in}
\caption{App downtime experienced by latency-sensitive edge apps during mobility. Details~\S\ref{sec:appendeix_implementatioin}.}
\vspace{-0.15in}
\label{fig:app_down_time_motivation}
\end{minipage} 
\vspace{-0.2 in}
    \hfill%
\begin{minipage}[t]{0.47\linewidth}
\includegraphics[width=\linewidth]{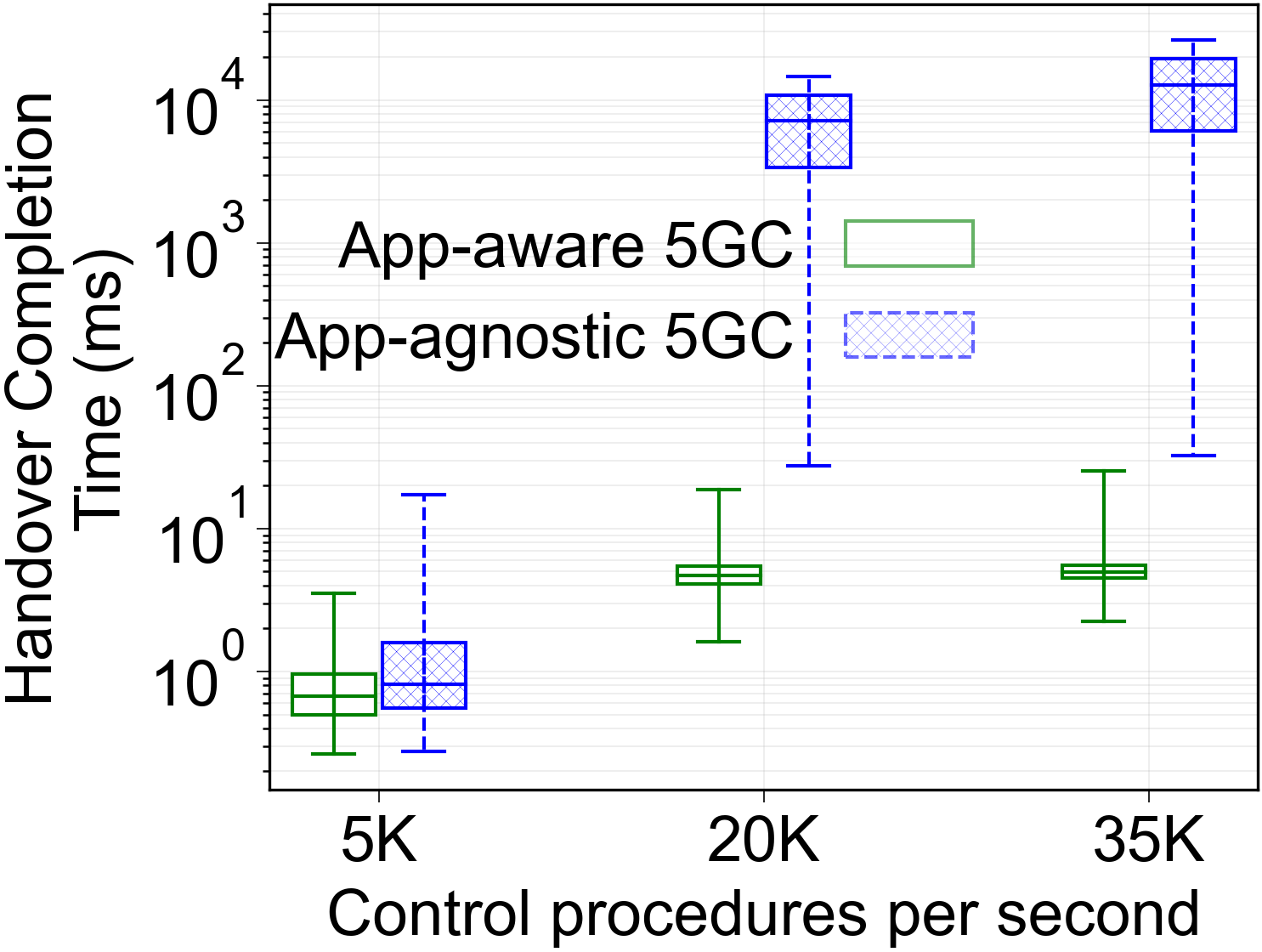}
\vspace{-0.3 in}
\caption{Handover completion time comparison when apps are served by different 5G cores.}
\label{fig:handover_motivation}
\end{minipage}%
\end{figure}

Figure~\ref{fig:app_down_time_motivation} shows the app downtime during an app handover (due to mobility) for three edge apps having different user-specific dynamic state sizes.\footnote{These results are obtained by interfacing the CarMap and EMP (\S\ref{sec:imp_modified_state_store}) with \Name and using our emulation framework ~(\S\ref{sec:edge_apps_simulation_framework}) for Mobile VR. We assume these apps require strict consistency \cite{tanenbaum}.} 
The Figure shows that app downtime for latency-sensitive apps can be up to 9$\times$ of their delay budget. 
Secondly, for all three apps, the downtime is larger than the delay budget for the respective apps. An app downtime exceeding the delay budget can degrade app performance, such as compromise user safety in the case of connected cars~\cite{zhang2021emp, ahmad_carmap} or cause users motion sickness when using immersive mobile VR~\cite{motion_sickness_kundu2021study}.


\subsection{Frequent Application Handover}
\label{sec:back:frequent_app_handover}
Prior studies suggest app handovers are likely to be frequent~\cite{frequent_handover_zhou20215g, frequent_migration_abouaomar2021deep,  driving_frequent_app_handoverr,mec_in_5g, mec_evolution_toward_5g_etsi, etsi2}.
A recent 5G measurement study~\cite{narayanan2020first} shows 30+ 5G handovers can occur in less than eight minutes duration. Similarly, another study shows that on average, 4G handovers can occur every 70 seconds in walking scenarios~\cite{dpcm}. When edge apps are deployed beside base stations, every network handover may trigger an app session migration. Even if edge apps are hosted at BS aggregation points, an app session migration may happen frequently for mobile users~\cite{ahmad2022idriving_latency_budget}. These app migrations can result in frequent app downtime, as discussed in \S\ref{sec:slow_app_handover}.

\section{Issues and Challenges}
\label{challenges}
In this section, we identify the key issues and challenges in enabling stateful mobile edge apps. We also highlight the impact of these issues through testbed experiments and qualitative discussion.

\subsection{Slow Application Session Migration}
\label{sec:slow_state_migration}
Most of the existing app session migration techniques are based on containers~\cite{live_migration_8000803,vm_container_migration_in_5g_mec, hardware_accelerated_live_migration, efficient_container_migration, docker_live_migratioin}, VMs~\cite{fast_service_migratioin_using_vm}, and user-space tools~\cite{criu,stateful_process_migration}. 
Table~\ref{tab:state_migration_techniques_summary} shows CarMap~\cite{ahmad_carmap} app downtime during an app handover when using these session migration techniques.  Table~\ref{tab:state_migration_techniques_summary} shows that container-based migration provides an app downtime >1.2\,s which is significantly higher than the delay budget~\cite{ahmad2022idriving_latency_budget} of latency-sensitive edge apps (e.g., delay budget\footnote{CarMap has a total delay budget of 100\,ms.
Processing stack/network communication takes $\approx$ 80\,ms/20\,ms
~\cite{ahmad2022idriving_latency_budget}.} for CarMap~\cite{ahmad_carmap} $\approx$20\,ms). VM-based migration is even slower. A widely used user-space tool CRIU~\cite{criu} checkpoints a running app as a collection of files and resumes it on the destination machine. Table~\ref{tab:state_migration_techniques_summary} shows that CRIU provides an app downtime >140\,ms, which is still above the delay budget of latency-sensitive apps. On the other hand, app-specific optimizations-based techniques are relatively fast~\cite{fast_check_point_recovery}, however, they require the app developer to identify the user's dynamic app state and the checkpointing occasions (preferably idle times) to avoid or reduce performance degradation \cite{wang2010behavioral, white2007scaling}.

\noindent\textbf{Challenge:} The generic session migration techniques, such as VM or container-based migration are slow. User space migration techniques, such as CRIU, are relatively fast but still exceed the delay budgets of many apps. Whereas app-specific techniques are not developer-friendly due to the development efforts involved in optimizing every app. We look at this problem from a different lens, asking the question: \textit{can the 5G network help expedite app session migration under mobility while requiring minimal app-specific changes from the developer?}

\begin{table}[tb]
\begin{center}
\begin{tabular}{| c | c | c | c |} 
\hline
\textbf{\makecell{App HO \\ Method}} & \textbf{\makecell{App \\downtime}} & \textbf{\makecell{App \\agnostic?}} & \textbf{\makecell{Meet App \\delay budget?}}\\
\hline
VM & > 4\,s & Yes& \textbf{No}\\
\hline
Container & > 1.2\,s & Yes& \textbf{No}\\
\hline
CRIU & > 140\,ms & Yes & \textbf{No}\\
\hline
\end{tabular}

\caption{Downtime for CarMap (network delay budget~\cite{ahmad2022idriving_latency_budget} $\approx$20\,ms) during an App handover (HO) with userspace tool CRIU~\cite{criu}, Container~\cite{docker_live_migratioin, vm_container_migration_in_5g_mec,live_migration_8000803, lxc_live_migration_edge} (Docker and LXC), and VMs~\cite{live_migration_8000803}. 
}
\label{tab:state_migration_techniques_summary}
\vspace{-0.3in}
\end{center}
\end{table}

\subsection{Reactive Application Session Migration}
\label{sec:challenges:reactive_app_session_migration}

When performing reactive app handover, the minimum app downtime is the maximum delay of app session migration and 5G handover (Figure~\ref{fig:challnges_serialized_ho}b). On the other hand, proactive session migration requires information about the target BS before a 5G handover starts. Our analysis of the 4G/5G radio traces,
 having \totalHandovers handover events shows on average 7.45 candidate BSs for a handover (range between 1-20). The latency-sensitive apps we explored have dynamic state sizes of $\approx$3\,MBs~(\S\ref{sec:eval:impact_on_carmap} and \S\ref{sec:eval:impact_on_emp}) and can have an update rate of 100 updates/s~\cite{zhang2017towards_online_game_updates_rate}. Proactively synchronizing app state on multiple BSs (on average 7.45) is highly inefficient in terms of bandwidth.

\noindent\textbf{Challenge:} Proactive app session migration on multiple BSs is not efficient. As multiple candidate BSs are available for a handover in which only one target BS is selected at the time of handover, \textit{it is challenging to determine the target BS sufficiently in advance and with reasonable accuracy} to enable proactive app session migration.

\subsection{Application-Agnostic 5G Handover}
\label{sec:app_agnostic_5g_handover}

To highlight how app-agnostic 5G handover events can impact the downtimes for edge apps, we conduct testbed experiments. Figure \ref{fig:handover_motivation} shows handover completion time for latency-sensitive apps in different control plane load scenarios; where the load corresponds to UEs serving different applications, such as AR/VR, connected and autonomous vehicles, and haptic feedback \cite{iot_in_5g_era_cisco, low_latency_erricson_industrial_iot, low_latency_erricson_industrial_iot_2}. The 5G control plane in Figure \ref{fig:handover_motivation} is serving a 12:88 ratio (projected IoT workload for 5G \cite{iot_traffic_distribution}) of latency-sensitive and latency-tolerant traffic.  We compare two schemes: (i) app-agnostic 5GC represents the existing control plane implementation where apps do not influence the control plane and (ii) app-aware 5GC represents an ideal 5G control plane that is aware of the delay budgets of the underlying apps running on UEs. When the load on the 5G control plane is low (i.e., 5K procedures\footnote{A procedure is a set of related request/response messages between a UE and cellular control plane, i.e., handover procedure.} per second), app-aware 5GC  provides a 1.2$\times$ improvement in median handover completion time over the app-agnostic 5GC. As the load increases on the control plane, we observe up to three orders of magnitude improvement in median handover completion times with app-aware 5GC. 
The load scenarios in Figure \ref{fig:handover_motivation} may be frequent in 5G as it aims to support massive machine-type communication with a connection density of up to 1 million devices per square km~\cite{5g_connection_density}.

\noindent\textbf{Challenge:} Edge sites are known to have limited resources \cite{trivedi2020sharing, edge_resouces, edge_resources_globe} therefore reserving control plane resources for every application class may not be efficient.\footnote{Network slices such as shown in Figure 5 of the \cite{ericsson_e2e_network_slices} reserves control plane for different application classes which may not be resource efficient in an edge-based cellular core deployment  \cite{iot_in_5g_era_cisco}.} In some cases, a shared control plane for different applications is inevitable when a UE connects to multiple network slices in 5G. A challenge here is to support latency-sensitive apps in an edge-based cellular core deployment with efficient resource utilization. In the existing 3GPP specifications \cite{3gpp_architecture, 3gpp_mme, 3gpp_handover}, the cellular control plane cannot differentiate between different application classes and it does not provide any interface/protocol for AF to influence the processing of the control plane.

\section{Design}
\label{sec:design}
To address the issues in \S\ref{challenges}, we design a new system architecture, \Name. \Name is a cross-layer design that (a) exposes radio network information (reported by the UE to the BS) through the MEC APIs~\cite{5g_mec_rni1} to edge apps so they can proactively perform application state migration, and (b) shares application latency requirements with the 5G control plane for speeding 5G handovers. 
\Name consists of three new modules: (i) a module for proactively predicting target BSs for mobile clients (\S\ref{sec:design:target_bs_predction}), (ii) an edge data store for stateful edge apps that can leverage proactive network information to speed up application state synchronization (\S\ref{sec:design_app_state_migration}), and (iii) app-aware control plane for speeding up 5G handover events (\S\ref{sec:app_impact_on_control_plane}). In this section, we first discuss our goals and assumptions in designing \Name (\S\ref{sec:goals_and_assumptions}), then provide an overview of our proposed architecture (\S\ref{sec:design:system_architecture}), before discussing the individual modules in detail.

\subsection{Goals and Assumptions}
\label{sec:goals_and_assumptions}

The key goal of \Name is to  \textit{\underline{reduce app downtime} for stateful edge apps when migrating application state due to client mobility}. At the same time, \Name aims to incur minimal latency and bandwidth overheads in non-mobility scenarios. We next make explicit our assumptions about the deployment and computational model before diving into our design.

\begin{itemize}[leftmargin=*]
    \item We assume edge deployment of the cellular core network which is the preferred deployment option for latency-sensitive apps \cite{iot_in_5g_era_cisco, ericsson_e2e_network_slices}.
   
   \item We assume a client-server model for apps. An app server hosted at the network edge provides services to mobile clients. A stateful app server maintains a per-user app state.
   
   
   \item We assume edge apps provide Read Your Writes consistency, which is \textit{the effect of a write operation by a process (user app) on data item x (user's app state) will always be seen by a successive read operation on x by the same process}. A system designed for Read Your Writes consistency can be easily tuned for apps that can tolerate relaxed consistency models \cite{tanenbaum}.

   \item We assume every edge site acts as an aggregation point for a set of BSs. Consequently, every 5G handover might not trigger an app state migration.

\end{itemize}

\subsection{\Name's System Architecture}
\label{sec:design:system_architecture}

\begin{figure}
\centering
\includegraphics[width=\columnwidth]{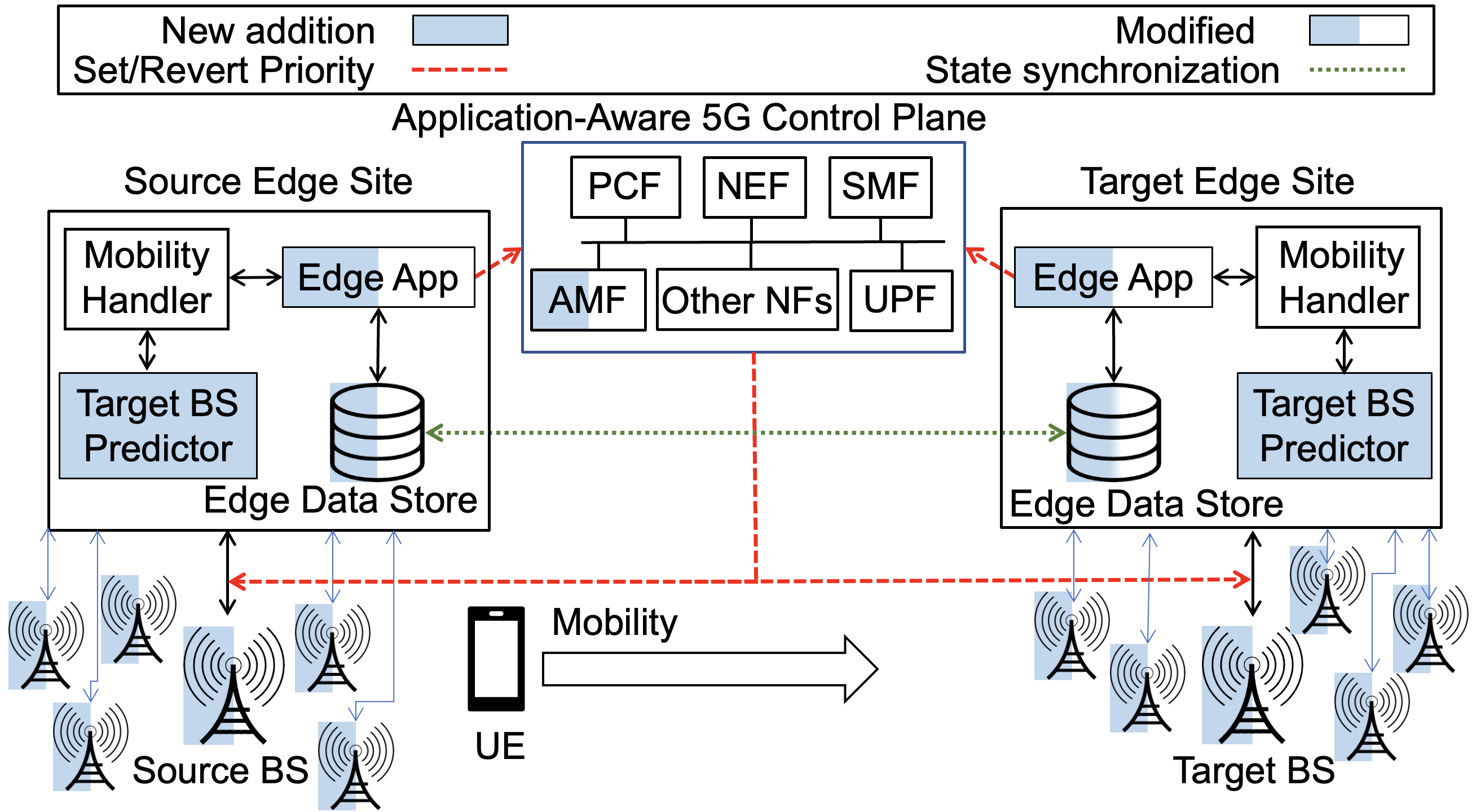}
\vspace{-0.3in}
\caption{\Name's system architecture.}
\label{fig:system_architecture_diagram}
\end{figure}

Figure~\ref{fig:system_architecture_diagram} provides an overview of \Name's system architecture. \Name argues to store the user's app state in our modified edge data store while the application logic is stateless.
To enable this, the modified data store exposes new APIs (Table~\ref{tab:api_in_edge_store}) to applications. The target BS predictor module receives real-time radio network information (RNI) exposed to applications in 5G \cite{5g_mec_rni1, mec_apis_combined} and predicts the destination BS in advance before an actual handover starts. The mobility handler module receives information about the target BS from the target BS predictor module and provides the target edge host and application instance information (such as IP address, port number, etc.) to the source app instance to facilitate proactive application state migration. The app-aware cellular control plane leverages the application-level information (such as latency needs) to provide priority processing for latency-sensitive edge apps. To enable this, the app-aware cellular control plane exposes new APIs~(Table~\ref{tab:api_for_app_influence_on_cpf}) to edge apps to prioritize their control plane processing. 

\subsection{Target BS Prediction}
\label{sec:design:target_bs_predction}


\Name introduces a novel approach to predict target BS well in advance of handover decisions, enabling proactive application state migration. Our multi-step pipeline tackles this as a multi-class classification problem with dynamic classes, where each BS represents a class and the set of neighboring BSs can change over time. The pipeline leverages real-time radio signal strength measurements from Radio Network Information (RNI) APIs~\cite{5g_mec_rni1}, utilizing Reference Signal Received Power (RSRP) and Reference Signal Received Quality (RSRQ). For high-accuracy predictions, we employ a stacked Long Short-Term Memory (LSTM) network. We chose LSTM for its ability to handle complex, non-linear mobile network data, capturing both short-term fluctuations and long-term patterns in rapidly changing 5G radio signals.

Our approach achieves 97\% prediction accuracy~(\S\ref{sec:eval:target_BS_prediction}) without extensive per-user data storage, outperforming other early target BS prediction schemes (typically <85\% accurate). The LSTM model, trained on real 4G/5G network traces from various mobility scenarios, forms the core of our prediction pipeline (Figure~\ref{fig:target_bs_prediction_architecture}):



\begin{figure}
\centering
\includegraphics[width=0.7\columnwidth]{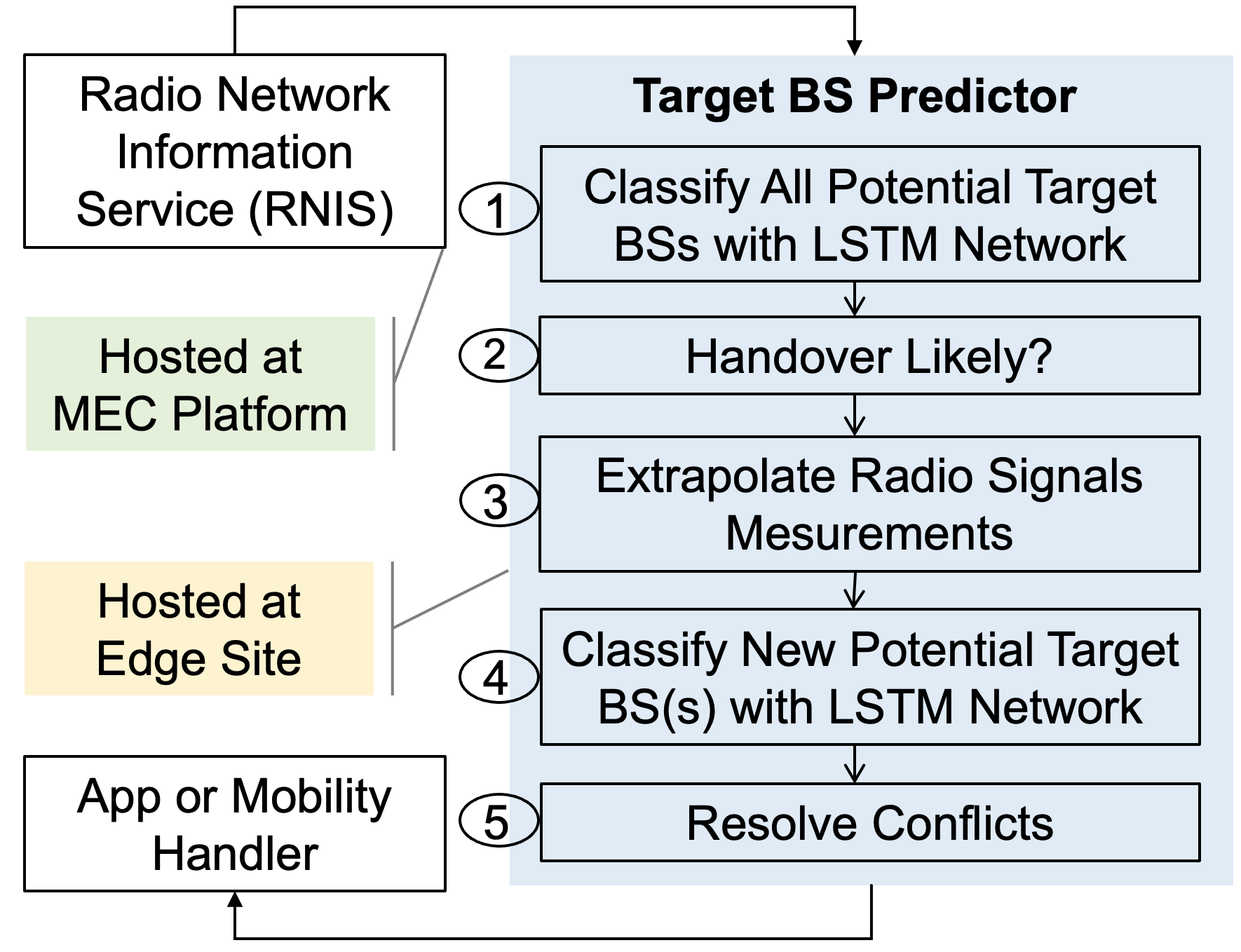}
\vspace{-0.1in}
\caption{\Name's target BS prediction.}
\vspace{-0.2in}
\label{fig:target_bs_prediction_architecture}
\end{figure}

\begin{enumerate}[leftmargin=*] 

\item \textbf{LSTM Network Prediction:} We apply an LSTM network~(\S\ref{appendix_lstm_binary_classifier}) to predict handover probabilities from the source BS to each neighboring BS with sufficient signal strength measurements. The network utilizes a feature vector (detailed in \S\ref{appendix_feature_vector}) for each BS pair. This vector comprises N absolute values and N relative values (compared to the source BS) of both RSRP and RSRQ measurements. The number N is configurable, with a default value of 6 set based on our evaluation results. This approach allows the LSTM to capture both the absolute signal strengths and their relative changes, enabling more accurate predictions of potential handovers.

\item \textbf{Handover Likelihood Assessment:} We flag a high likelihood of handover when the predicted handover probability to any target BS exceeds a learned threshold, let's say P\footnote{We learned the value of P = 0.37 that maximizes the F1-score of the binary LSTM classifier.}. 

\item \textbf{Early Prediction Handling:} For cases where the actual target BS has fewer than N samples (approximately 5\% of cases in our radio traces), we perform zero-order extrapolation of RSRP and RSRQ to complete the required window size. 

\item \textbf{Extrapolated Target BSs Prediction:} We apply the LSTM classifier to new source-target BS pairs as they become available. 

\item \textbf{Conflict Resolution:} We determine the top-x BSs from the pool of candidates based on handover probabilities assigned by the LSTM network. The value of x is configurable (default = 1) to accommodate specific edge application requirements. \end{enumerate}


\subsection{Two-Step App State Synchronization}
\label{sec:design_app_state_migration}

In this section, we discuss our modifications to the existing data stores to provide two-step state synchronization based on mobility hints. 

\noindent{\textbf{Modification to Data Store:}} Our modifications to the data store include keeping an additional meta-state with every key to track whether a state object is updated after the previous synchronization event and sorting state objects based on their update rates. The additional per-key meta-state in the data store at the source edge host includes:
\begin{itemize}[leftmargin=*]
    \item \textit{First update time}: The time the key was created in the data store.

    \item \textit{Last update time}: The time when the key was last time updated in the data store.

    \item \textit{Updates count}: The total number of key updates.

    \item \textit{Last sync time}: The time when the key was last synced with the data store instance at the target edge host. To avoid any conflict when an update and sync operation are performed in parallel, the external data store records the time before the sync process starts for a key (say T) and updates T as the last sync time when the sync operation completes.
\end{itemize}

The meta-state gives useful information about (i) the  average key update rate and (ii) whether the key is updated after the last sync. We leverage the per-key meta-state to define additional APIs (in Table \ref{tab:api_in_edge_store}) in the modified data store.

\noindent{\textbf{Background State Synchronization:}} An edge app registers itself with the mobility handler module (Figure~\ref{fig:system_architecture_diagram}) to receive important network events, such as mobility hint and mobility handover. The mobility handler module receives target BS information from the target BS prediction module and provides destination edge host and app instance information (i.e., IP address, port number) to the source edge app.
On reception of a mobility hint, the source edge app calls \textit{BackgroundSync} API exposed by our modified data store.
Call to the \textit{BackgroundSync} API initiates the syncing of the user's app state with the target data store instance by synchronizing the state object that has the lowest update rate and was updated after the last sync event (update time higher than sync time).
The background syncing blocks updates to the key only when getting a copy of the state. The migration and restoration in the target data store instance is a non-blocking operation. As read operations are very fast in edge-suitable data stores (such as \redis and \memcache), this allows the app to process users' requests with minimal overhead during background state syncing.

\noindent{\textbf{Blocking State Synchronization:}} 
On a 5G handover signal from the mobility handler module, the source edge app calls \textit{BlockingSync} API in Table \ref{tab:api_in_edge_store} by providing it with a callback function. This API synchronizes the rest of the state that is either not synced or updated after the last sync. In the best case scenario, all the user's app state is migrated during the background syncing and the callback function returns right after the call to the \textit{BlockingSync} API. In the worst case, all the user's app state needs to be synced due to post-sync updates. The duration from call to  \textit{BlockingSync} API to the callback call is the blocking state migration time---also the minimum app downtime during an app handover. If the third parameter (retain) in \textit{BlockingSync} API is true, the source data store instance will not delete the user state otherwise it will be deleted after the migration completes.


\begin{table}
\centering
\begin{tabular}{| c |} 
\hline
    \textbf{New APIs in Modified Data Store}\\
    \hline
    \textit{BackgroundSync (userID, destHost)}\\
    \hline
    \textit{BlockingSync (userID, callBackFunc, retain)}\\
    \hline
\end{tabular}
\caption{New APIs for latency-sensitive edge apps in the modified data store tailored to the needs of latency-sensitive mobile applications.}
\label{tab:api_in_edge_store}
\vspace{-0.25in}
\end{table}

\begin{table}
\centering
\begin{tabular}[\columnwidth]{| c| c |} 
    \hline
    NF & \textbf{New APIs in App-Aware 5GC for AF}\\
    \hline
    AMF & \textit{Set\_Priority (IpAddress, priority)}\\
    \hline
    AMF & \textit{Revert\_Priority (IpAddress)}\\
    \hline
    RAN & \textit{Set\_Priority (RAN UE NG-AP ID, priority)}\\
    \hline
    RAN & \textit{Revert\_Priority (RAN UE NG-AP ID)}\\
    \hline
\end{tabular}
\caption{New APIs available in \Name for an AF to influence control plane operations at the RAN/AMF.}
\label{tab:api_for_app_influence_on_cpf}
\vspace{-0.25in}
\end{table}

\subsection{App-Aware Control Plane Processing}
\label{sec:app_impact_on_control_plane}
In this section, we discuss how an AMF can differentiate between different types of UEs. We also describe how an AMF leverages this information to perform priority processing for the UEs running a latency-sensitive app session.

\noindent{\textbf{5G Services Classes:}} 5G networks are aiming to support three broad classes of services, (i) enhanced Mobile Broad Band (eMBB), (ii) Ultra-Reliable and Low Latency Communication (URLLC), and (iii) delay-tolerant low-complexity Massive IoT (MIoT) \cite{3gpp_architecture, latency_nbiot_catm, iot_in_5g_era_cisco,critical_iot, 3gpp_evaluation_mmtc}.

\noindent{\textbf{UE Classes at AMF:}}
 An AMF can differentiate between the low-complexity (MIoT) and other (eMBB \& URLLC) devices from the information provided by RAN or UE.\footnote{UE context at AMF (table 5.2.2.2.2-1 of the 3GPP document~\cite{3gpp_23_502}) contains information if a UE is a low-complexity device such NB-IoT, LTE-M or RedCap (reduced capability device). Section 1.8 of the 3GPP document~\cite{3gpp_124_008_nas_low_priority_info} suggests a UE can explicitly inform the core network that it is a low-priority device.} An AF, on behalf of app~\cite{mec_in_5g}, can inform which eMBB or URLLC device is to be treated normally or with high priority by the control plane. 
In the context of delay tolerance, our modified AMF can differentiate between the following classes of UEs:

\begin{enumerate}[leftmargin=*]
    \item {\lowPriorityIot:} These are delay-tolerant low-complexity devices such as those used for smart cities, e.g., smart meters. A delay of up to 10\,s is acceptable for these UEs \cite{latency_nbiot_catm, iot_in_5g_era_cisco, 3gpp_evaluation_mmtc}. \Name's AMF is configured to have \textit{lowest priority} for such UEs.
    \item {\mediumPriorityIot:} 
    Includes the rest of devices not included in \lowPriorityIot, such as eMBB and URLLC.
    Such UEs are typically used in agriculture, education, media production, etc \cite{iot_in_5g_era_cisco, cellular_iot_ericsson}. Voice services can also be included in this class of UEs. These UEs have \textit{medium priority} in \Name. 
    \item {\highPriorityIot:} These are devices from the \mediumPriorityIot having an active latency-sensitive data session with an edge app. These UEs have deadlines in the order of a few ms \cite{critical_iot, iot_in_5g_era_cisco}. Autonomous vehicles \cite{self_driving_cars_lin2018architectural} and VR headsets \cite{vr-latency} are key examples of such UEs. The server side of a latency-sensitive edge app adds a UE to this class when an eMBB or URLLC device starts a latency-sensitive data session with it. These UEs have \textit{highest priority} in \Name.
\end{enumerate}

\noindent\textbf{{Priority Processing for Latency-Sensitive Apps:}}
We leverage the UE class information at the AMF to perform priority processing for latency-sensitive apps at RAN and AMF. Radio Resource Control (RRC) layer~\cite{3gpp_architecture} provides a control plane signaling protocol between a UE and RAN. We maintain three separate priority queues for the RRC messages at the BS. The Next Generation Application Protocol (NG-AP) provides the control plane signaling protocol between the RAN and AMF. We maintain three priority queues at the BS for the NG-AP messages from the AMF. Similarly, we maintain three queues at the AMF from the control messages received from the BS. UEs at the RAN and AMF are identified with their unique IDs, \textit{RAN UE NGAP ID} and \textit{AMF UE NGAP ID}, respectively.
 The priority levels are configurable in our design and can be updated by network operators based on their preferences.\footnote{The control processing threads at the BS and AMF by default process 6/10, 3/10, and 1/10 control messages from \highPriorityIot, \mediumPriorityIot, and \lowPriorityIot queues, respectively.} If a high-priority queue is empty, we process messages from one level lower priority queue. Our design requires no changes at the UE. 
 In section \S\ref{appendix_d}, we provide a detailed discussion of the sequence of steps involved when an edge app calls the APIs in Table~\ref{tab:api_for_app_influence_on_cpf} to prioritize their processing.

\section{Implementation}
\label{sec:implementation}

\subsection{Target BS Prediction}
\label{sec:impl_target_bs_prediction}
We implement the LSTM network for the target BS prediction in Python on top of the \textit{Keras version 2.9.0}. The LSTM network has a total of 8 layers and 15,923 trainable parameters. 
Our data set for training the LSTM network consists of radio network traces collected in high-speed trains~\cite{hst_sigcomm_dataset}, driving tests~\cite{incp_dataset}, and other miscellaneous mobility scenarios~\cite{mobile_insight_logs}. The data set consists of a total of \totalHandovers handover events. The radio traces in the data set are sampled at a $\approx$50\,ms interval.

\subsection{Modified External Data Store}
\label{sec:imp_modified_state_store}
We implement all our modifications (as discussed in \S\ref{sec:design_app_state_migration}) in the \redis~\cite{redis_store} and \memcache~\cite{memcached_store} data stores. We implement our new APIs in Table \ref{tab:api_in_edge_store} with approximately 1,392 lines of code changes in the open-source code~\cite{redis_code} for \redis. For \memcache, we made our changes on top of its common client APIs (get, set, delete, etc.).

We used three versions of \redis clients, each one in C++~\cite{redis_cpp}, Java~\cite{redis_java}, and Python~\cite{redis_python}. We also implement a Python version of the modified \memcache data store on top of the \textit{pymemcache 4.0.0} Python package. We interfaced our modified data stores with the following apps:


\noindent\textbf{CarMap:} We modify the open-source C++ code-base of CarMap \cite{carmap_gitub, ahmad_carmap} to use the C++ version of our modified \redis data store for application state storage and updates. Our changes to CarMap comprise of $\approx$477 lines of code.  

\noindent\textbf{EMP:} We modify the open-source Java code-base of EMP \cite{emp_gitub, zhang2021emp} to use the Java version of our modified \redis data store. Our changes to EMP consist of $\approx$350 lines of code. 

\subsection{Edge App Simulation Framework}
\label{sec:edge_apps_simulation_framework}
State properties across edge apps differ and it's impractical to update many apps due to limited options. Therefore, we created a Python-based simulation framework for better experimental coverage. It allows users to configure their app state properties (i.e., state size,  dynamic state percentage, state update rate) on the server side and generates app traffic on the client side. The framework interfaces with both modified \redis and \memcache data stores. The simulation framework consists of $\approx$1147 lines of code.

\subsection{App-Aware Control Plane Processing}
\label{sec:imp_app_aware_control_plane_processing}
We have modified the open-source C/C++ code-base of a recently proposed cellular control plan~\cite{neutrino_gitub, neutrino_ton} to implement app-aware control plane processing (\S\ref{sec:app_impact_on_control_plane}). 
We implement the RAN-related APIs of Table \ref{tab:api_for_app_influence_on_cpf} in the control traffic generator (emulating both UE and BS) of \cite{neutrino_gitub}. Similarly, we implement the AMF-related APIs of Table \ref{tab:api_for_app_influence_on_cpf} in the control plane function (AMF + SMF) of \cite{neutrino_gitub}. 
Our changes to \cite{neutrino_gitub} consist of the addition of $\approx$967 lines of code.

\section{Evaluation}
\label{sec:evaluation}

In this section, we show the following:

\begin{enumerate}[leftmargin=*]
    \setlength{\itemsep}{0pt}
    \setlength{\parskip}{0pt}
    
    \item \textbf{Improvement in Downtime for Real Edge Apps:} \Name reduces app downtime during an app handover by up to $14\times$ and $15.4\times$ for CarMap and EMP, respectively.
    
    \item \textbf{Impact of Proactive App State Migration:} \Name reduces blocking app state migration time
    by more than $2.5\times$ for both CarMap and EMP.
    
    \item \textbf{Impact of App-Aware Control Plane Processing:} \Name reduces 5G handover completion time by up to 3 orders of magnitude for latency-sensitive apps.
    
    \item \textbf{Comprehensive Apps Coverage:} Our sensitivity analysis using our simulation framework~(\S\ref{sec:implementation}) shows \Name can reduce application downtime for a wide variety of applications.
\end{enumerate}

\subsection{Settings and Methodology}
\label{sec:app_handover_setting_and_methodology}
\begin{figure}[]
\centering
\includegraphics[width=\columnwidth]{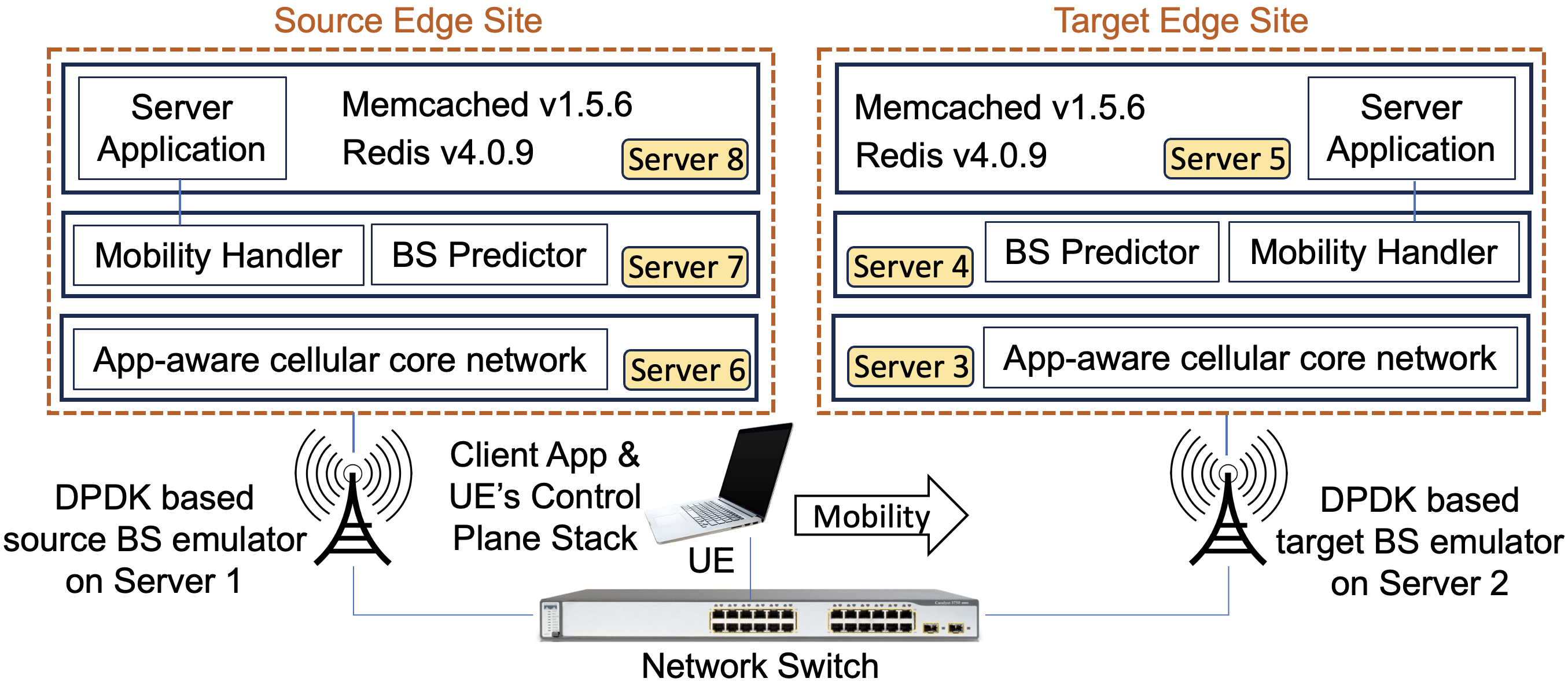}
\vspace{-0.3in}
\caption{Testbed for experiments.}
\label{fig:eval:testbed}
\vspace{-0.2 in}
\end{figure}

We evaluate \Name on a testbed comprising one client laptop and eight dual-socket servers, each hosting different components (Figure~\ref{fig:eval:testbed}). Both source and destination sites consisted of a BS emulator, a cellular core network, edge applications, and a mobility handler with a target BS predictor module. The servers operate on an Intel Xeon(R) Gold 5220 CPU @ 2.20\,GHz and have 128\,GB memory, while the laptop has 20\,GB RAM, 512\,GB SSD, and a GeForce GTX 1060 6\,GB GPU. 
To ensure statistical significance and capture variability, we repeated each experiment a minimum of 100 times for all plots displaying distributional data. 

\subsection{Overall Improvement with Real Apps}
\label{sec:eval:combined_impact}

\begin{figure}[]
\centering
\includegraphics[width=0.85\columnwidth]{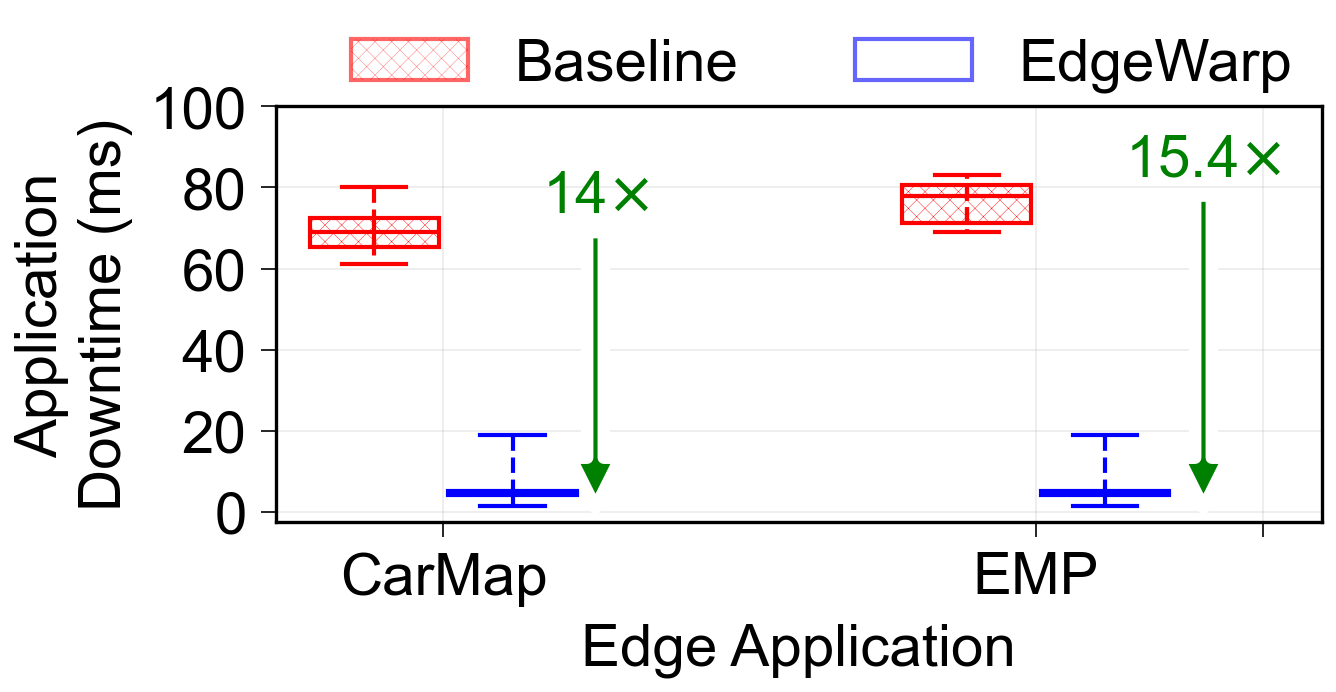}
\vspace{-0.2in}
\caption{Overall improvement in app downtime during an app handover for CarMap~\cite{ahmad_carmap} and EMP~\cite{zhang2021emp}.}
\vspace{-0.2in}
\label{fig:eval:oveerall_app_downtime}
\end{figure}

In this section, we show improvement in app downtime with both proactive application state migration and app-aware control plane processing of \Name. Figure \ref{fig:eval:oveerall_app_downtime} shows app downtime during an app handover for CarMap~\cite{ahmad_carmap} and EMP~\cite{zhang2021emp}. We use the default configuration for both CarMap and EMP.\footnote{Row 1 of Table~\ref{table:eval:carmap_state_migration_time} and Table~\ref{table:eval:emp_state_migration_time} shows default configuration for CarMap and EMP, respectively.} In Figure \ref{fig:eval:oveerall_app_downtime}, ``Baseline'' represents coordinated but reactive app handover (i.e., representing the coordinated handover as shown in 
Figure \ref{fig:challnges_serialized_ho}b) and ``\sysname'' represents the modified application version with our proposed design (Figure \ref{fig:challnges_serialized_ho}d). Figure~\ref{fig:eval:oveerall_app_downtime} shows  $14\times$ and $15.4\times$ improvement in median app downtime for CarMap and EMP, respectively, when using \Name. In the following sections, we highlight the contribution of different modules to the overall improvement in app downtime.


\begin{figure}[]
\centering
\includegraphics[width=\columnwidth]{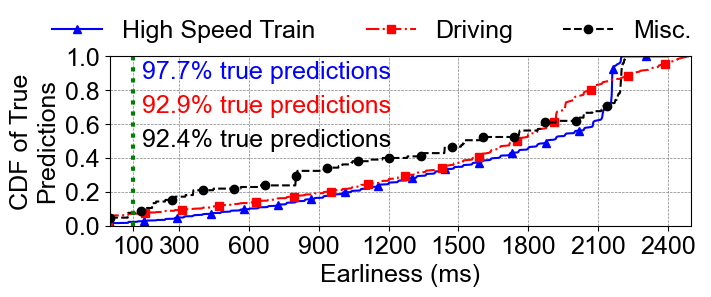}
    \vspace{-0.35 in}
    \caption{Accuracy of the early target BS prediction on three real data sets. The annotated statistics show the accuracy of predicting target BS 100 ms ahead of handover.
    }
    \vspace{-0.2 in}
    \label{fig:eval:cdf_target_bs_prediction}
\end{figure}

\subsection{Target BS Prediction with Real Traces}
\label{sec:eval:target_BS_prediction}

Figure~\ref{fig:eval:cdf_target_bs_prediction} shows the earliest time to predict a target BS with the real 5G/4G radio network traces collected in different mobility scenarios. The figure shows \drivingTestAccuracy, \highSpeedTrainAccuracy and \miscAccuracy true predictions 100\,ms before an actual handover with traces collected during driving~\cite{incp_dataset}, in high-speed trains~\cite{li2020beyond}, and miscellaneous mobility scenarios~\cite{mobile_insight_logs}, respectively. The figure shows that with a 100\,ms advanced mobility hint (said differently, predicting the target BS 100~ms before the handover), on average \averageAccuracy true handover predictions are possible. The theoretical upper limit of true predictions is \highSpeedTrainMaxAccuracy, \drivingTestMaxAccuracy, and \miscMaxAccuracy in high-speed trains, during driving, and miscellaneous mobility scenarios, respectively. The percentage is computed as the proportion of times when at least one radio sample is received for the true target BS before the handover.
A mobility hint received 100\,ms in advance is enough for the edge apps we tested---both CarMap (a localization app) and EMP (a safety app) require less than 100\,ms to provide less than 10\,ms blocking state migration time with default configurations (row 1 of the table~\ref{table:eval:carmap_state_migration_time} \& \ref{table:eval:emp_state_migration_time}). Other safety and localization apps with similar state properties are also expected to perform similarly to CarMap and EMP.

We also evaluate the inference time for the BS prediction pipeline. For a batch of 50 and 500 users, the median inference times are 8.7 ms and 9.5 ms, respectively
(\S\ref{sec:appendix_inference_time}).




\subsection{Impact of Proactive State Migration on Real Applications}
\label{sec:state_migration_delay}
This section assesses the efficacy of \Name's two-stage state synchronization in real edge apps using \redis store.

\subsubsection{Improvement in CarMap with \Name}
\label{sec:eval:impact_on_carmap}

\newcommand\tikzmark[2]{%
\tikz[remember picture,overlay] 
\node[inner sep=0pt,outer sep=2pt] (#1){#2};%
}

\newcommand\link[2]{%
\begin{tikzpicture}[remember picture, overlay, >=stealth, shorten >= 1pt]
  \draw[->] (#1.east) to  (#2.west);
\end{tikzpicture}%
}

\begin{table}
\centering
\begin{tabular}{c c c c}
\hline\hline 
App State & Dynamic & \multicolumn{2}{c}{Migration Time (ms)}  \\
\cline{3-4}
Size (MBs) & State (\%) & \multicolumn{1}{c}{Baseline} & \multicolumn{1}{c}{\Name} \\
\hline 
2.3 & 4.1 & \tikzmark{a}{82.4} &  \tikzmark{1}{6.8}  \\ 
4.3 & 47.1 & \tikzmark{b}{134.2} & \tikzmark{2}{45.5}  \\
5.4 & 57.3 & \tikzmark{c}{158.9} & \tikzmark{3}{62.8}  \\
\hline 
\end{tabular}
\link{a}{1}
\link{b}{2}
\link{c}{3}
\caption{Reduction in CarMap's average \underline{blocking state migration} time with \Name's modified \redis data store. We vary the frame upload interval at the vehicle to experiment with different state sizes.}
\label{table:eval:carmap_state_migration_time} 
\vspace{-0.25in}
\end{table}

In Table \ref{table:eval:carmap_state_migration_time}, we compare the performance of CarMap during an app handover in terms of blocking application state migration time---the time from the start of a 5G handover to the completion of state migration. 
We compare two versions of CarMap: (i) CarMap Baseline represents existing systems where application state migration starts with a 5G handover and (ii) CarMap with \Name represents the case when the application state is stored in our modified \redis data store and two-step state synchronization is performed. Here the two-step state synchronization is performed with a mobility hint of \mobilityHintWindow that results on average in \averageAccuracy correct target BS predictions (Figure~\ref{fig:eval:cdf_target_bs_prediction}). Table \ref{table:eval:carmap_state_migration_time} shows that with \Name, the application blocking state migration time drops to 6.8\,ms and 62.8\,ms from 82.3\,ms and 158.9\,ms when the application dynamic state varies from 4.1\% to 57.3\%, respectively. 
CarMap gets the benefits of proactive application state migration in \averageAccuracy cases and performs like the baseline in 3\% cases. 

\begin{figure}[]
\centering
\includegraphics[width=\columnwidth]{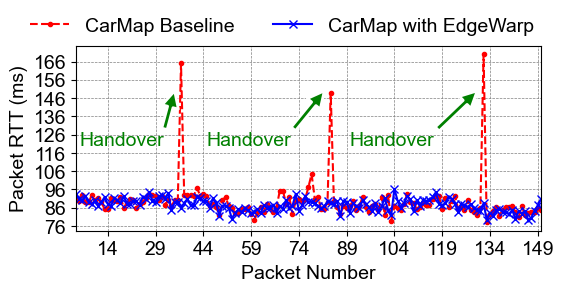}
\vspace{-0.35in}
\caption{Improvement in message RTT during an app handover when using CarMap with \Name as compared to CarMap Baseline. Settings row 1 of table~\ref{table:eval:carmap_state_migration_time}.}
\vspace{-0.15in}
\label{fig:eval_carmap_rtt}
\end{figure}

Figure \ref{fig:eval_carmap_rtt} shows message RTT measured on the vehicle when three app handovers are performed during an experiment. The figure shows $\approx$2$\times$ increase in RTT during an app handover when the vehicle client is served with CarMap Baseline. In contrast, CarMap with \Name does not see any significant change in message RTT during an app handover. This improvement is primarily attributed to the mobility hint-based two-step state synchronization provided in the modified \redis data store.



\subsubsection{Improvement in EMP with \Name}
\label{sec:eval:impact_on_emp}

\begin{table}
\centering
\begin{tabular}{c c c c} 
\hline\hline
App State & Dynamic & \multicolumn{2}{c}{Migration Time (ms)}\\
\cline{3-4}
Size (MBs) & State (\%) & \multicolumn{1}{c}{Baseline} & \multicolumn{1}{c}{\Name}  \\
\hline 
0.46 & 100 & \tikzmark{d}{77} & \tikzmark{4}{0.04}  \\ 
0.71 & 100 & \tikzmark{e}{89} & \tikzmark{5}{28}  \\
0.95 & 100 & \tikzmark{f}{145} & \tikzmark{6}{59} \\
\hline 
\end{tabular}
\link{d}{4}
\link{e}{5}
\link{f}{6}
\caption{Reduction in EMP's average \underline{blocking state migration time} when using \Name's modified data store. We vary frames per vehicle at the edge server to experiment with different state sizes.} 
\label{table:eval:emp_state_migration_time} 
\vspace{-0.35in}
\end{table}

In Table \ref{table:eval:emp_state_migration_time}, we compare the performance of EMP baseline and EMP integrated with the modified \redis data store during an app handover. The table shows that the blocking application state migration time drops to $\approx$ 0~ms when using \Name's modified data store with app state size $\approx$ 0.46\,MBs. EMP with \Name shows 3.2$\times$ and 2.5$\times$ improvement over the baseline with app state size 0.71\,MBs and 0.95\,MBs, respectively. The app state properties depend on the application logic. EMP has a relatively smaller state size as compared to CarMap but provides higher blocking state migration time because all the application state is dynamic. 

Our evaluation also shows more than 3$\times$ improvement in message RTT measured on the vehicle under mobility for EMP when using \Name
~(further details in \S\ref{appendix_rtt_emp}).

\subsection{State Migration Sensitivity Analysis}
\label{eval:factor_analysis}
In this section, we show the impact of varying mobility hints on blocking state migration time for CarMap. As real apps provide limited options to experiment with different state properties, we also show the impact of varying dynamic state sizes with our simulation framework~(\S\ref{sec:impl_target_bs_prediction}).

\subsubsection{Impact of mobility hint}
\label{sec:eval:impact_of_mobility_hint}
\begin{figure}[]
\centering
\includegraphics[width=\columnwidth]{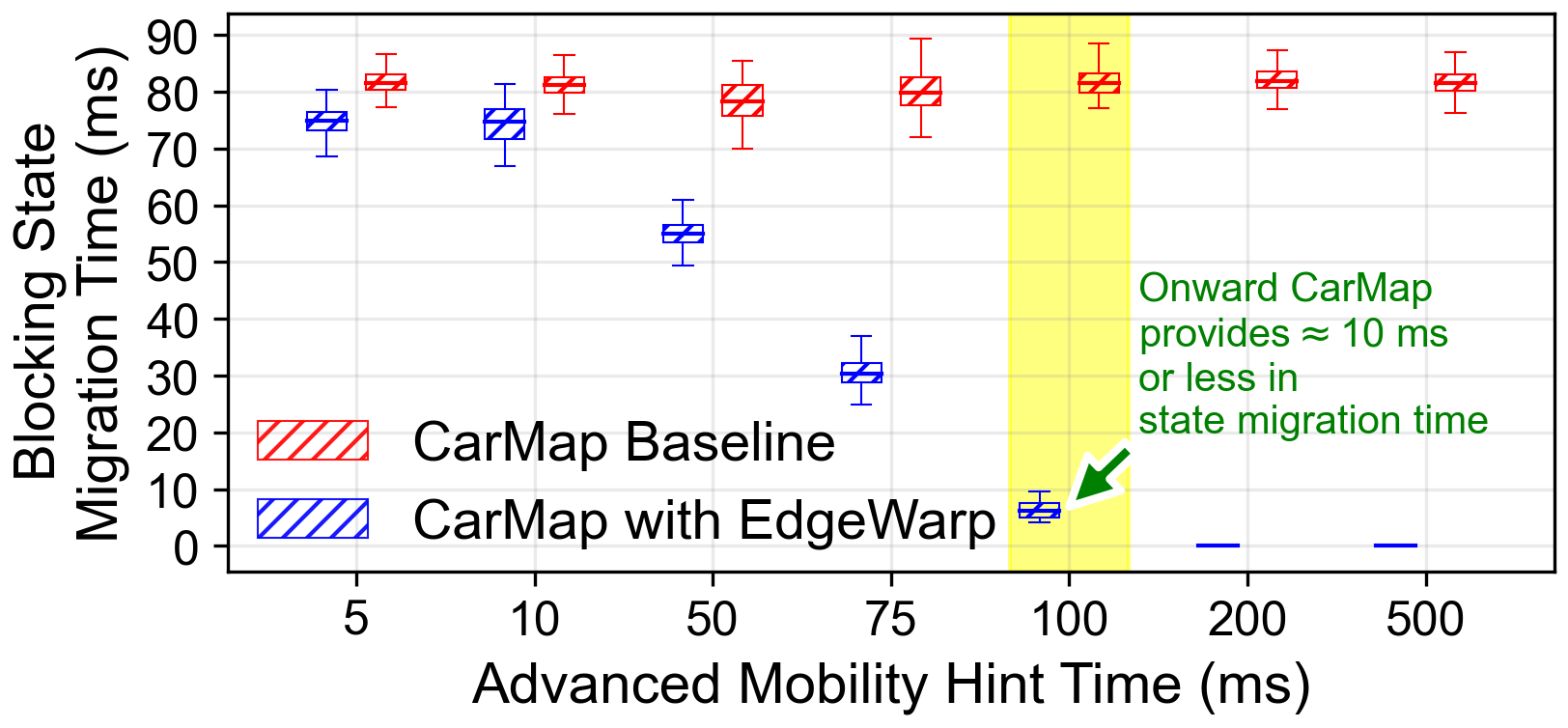}
\vspace{-0.3in}
\caption{Impact of early mobility hint on CarMap's blocking state migration time. Frame upload interval = 1\,s. Delay budget $\approx$20\,ms~\cite{ahmad2022idriving_latency_budget}.}
\vspace{-0.2in}
\label{fig:migration_time_vs_mobility_hint}
\end{figure}

The advanced mobility hint varies across handovers (as shown in Figure \ref{fig:eval:cdf_target_bs_prediction}) due to changing radio signal strength. In Figure \ref{fig:migration_time_vs_mobility_hint}, we quantify the impact of different advanced mobility hints on CarMap's 
blocking state migration time. CarMap baseline performs very close to CarMap with \Name when the advanced mobility hint is under 10\,ms. CarMap with \Name performs 1.5$\times$ better than CarMap baseline in median blocking state migration time with an advanced mobility hint of 50\,ms. When the advanced mobility hint is greater than 100\,ms, CarMap with \Name provides a median blocking state migration time close to zero as compared to 80\,ms for CarMap baseline. The figure shows an advanced mobility hint of 100\,ms reduces CarMap blocking state migration time well below its network delay budget of 20~ms~\cite{ahmad_carmap}.

\subsubsection{Impact of varying dynamic App state size}
\label{sec:eval:impact_of_dynamic_app_state_
size}

\begin{figure}[]
\centering
\includegraphics[width=\columnwidth]{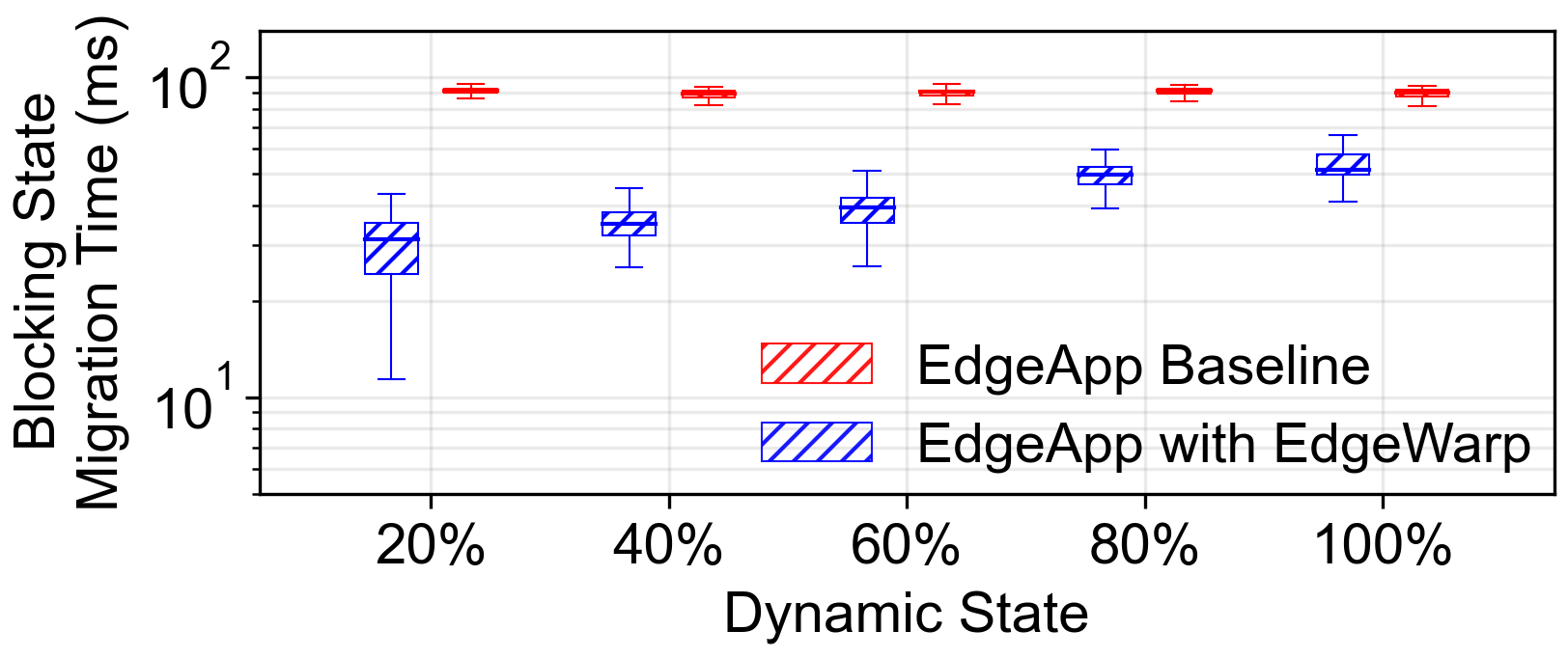}
\vspace{-0.3in}
\caption{\Name's application downtime with varying dynamic state size. The dynamic state update rate = 50 updates/s. The Y-axis is on a log scale.}
\vspace{-0.1in}
\label{fig:migration_time_dynamic_size_impact_memcache}
\end{figure}

To understand \Name's impact on a variety of apps, we simulate applications having different state properties with our simulation framework. The simulated apps store and perform proactive app state migration with our modified \memcache data store (\S\ref{sec:imp_modified_state_store}). In Figure \ref{fig:migration_time_dynamic_size_impact_memcache}, we compare the blocking state migration time for the simulated app (baseline) and with \Name. 
We vary the app dynamic state from 20\% to 100\% with a total state size of 3\,MBs (the observed dynamic state size for real edge apps~, as shown in table~\ref{table:eval:carmap_state_migration_time} \& \ref{table:eval:emp_state_migration_time}). We set the update rate of every dynamic state object to 50 updates/s to simulate high-frequency updates~\cite{zhang2017towards_online_game_updates_rate}.
With 20\% dynamic state, \Name performs 2.9$\times$ better in median blocking state migration time as compared to the baseline. The improvement of \Name over the baseline reduces as the dynamic state percentage increases. With a 100\% dynamic state, the performance of \Name is still 1.7$\times$ better as compared to the baseline. This is due to \Name's capability of proactively migrating some or all the app state following advanced mobility hints.


\subsection{App-Aware Control Plane Impact}
\label{sec:app_aware_control_plane_processing}
In this section, we compare the performance of \Name with \Old---a DPDK-based Open-Air Interface  implementation of 5G control plane~\cite{neutrino}
where the AMF is not aware of the user app.

\begin{table}
\centering
\begin{tabular}{l c c c c} 
\hline\hline 
Control Procedures/second & 20K & 25K & 30K & 35K \\ [0.5ex] 
\hline
Without \Name & 23\% & 30\% & 36\% & 42\% \\ 
With \Name & 0\% & 0\% & 0\% & 0\% \\
\hline
\end{tabular}
\caption{The \% of missed deadlines for CarMap, during a handover by \Old which drops to zero with \Name. Load represents control procedures per second. Lower \% is better.}
\label{table:control_plane_impact_on_apps}
\vspace{-0.33in}
\end{table}

\subsubsection{Impact on app performance}
\label{sec:app_impact_on_app_performance}

It is projected~\cite{iot_traffic_distribution} that 5G IoT will constitute 60\% of low-priority apps (\lowPriorityIot), 28\% for medium-priority apps (\mediumPriorityIot), and 12\% for high-priority apps (\highPriorityIot).  Latency-sensitive connected and autonomous vehicles, such as CarMap, have a network delay budget of $\approx$20\,ms~\cite{ahmad2022idriving_latency_budget}. Table \ref{table:control_plane_impact_on_apps} illustrates the percentage of responses that missed deadlines\footnote{The missed app deadlines are solely attributed to the 5G handover. State migration is not a factor in this experiment.} for the CarMap app when served by \Old.
The table indicates a 19\% increase in missed deadlines for CarMap as the load increases from 20K to 35K (a moderate load range~\cite{mmlite, neutrino}) when served with \Old. In contrast, when CarMap is served with \Name, the percentage of missed deadlines drops to zero across all load scenarios.
The performance gain for CarMap in Table~\ref{table:control_plane_impact_on_apps} is due to more than two orders of magnitude reduction in \handover completion time when using \Name (details in \S\ref{sec:appendix_control_plan_bursty_traffic}).

\section{Related Work}
\label{sec:related_work}

\noindent\textbf{Target BS Prediction:}
In prior work~\cite{bs_prediction_radio_link_characteristics, bs_prediction_ho_history, bs_prediction_low_accuracy_1}, target BS prediction is performed using support vector machines, handover history, and LTE event patterns. However, these studies focused on pre-5G networks, which differ from 5G with mmWave having limited range and more susceptible to physical obstacles~\cite{5g_mmWave}. Recent 5G-based approaches employed deep neural networks~\cite{bs_prediction_low_accuracy_2}, historical handovers~\cite{bs_prediction_low_accuracy_3}, and LSTM networks~\cite{bs_prediction_low_accuracy_4}. Notably, all these methods lack early prediction timings and provide relatively low prediction accuracy (<85\%).

\noindent\textbf{Resilience, Placement, and Load Balancing at the Edge:} 
CESSNA \cite{harchol2018cessna} uses message logging and periodic state check-pointing for consistent recovery in edge failure and client mobility cases. Portkey ~\cite{noor2021portkey} designs a key-value store adapting data placement per client mobility and data patterns, using passive profiling and active probing. EdgeBalance~\cite{zhang2020edgebalance} achieves load balancing in edge apps through CPU load prediction and an adaptive server load distribution. All these proposals can be used with our proposed design to further improve the edge apps' performance.


\noindent\textbf{Mobility Impact on Application Performance:}
A recent work~\cite{hassan2022vivisecting} quantifies the impact of mobility on app performance, power consumption, and signaling overheads. This work designs a \handover prediction module \textit{Prognos} and demonstrates its ability to improve real app performance. This work does not consider the problem of application session migration during mobility. In addition, this work does not speed up 5G handovers for latency-sensitive edge apps.

\noindent\textbf{Reducing Cellular Control Plane Latency:}
Recent cellular control plane designs~\cite{sigcomm22_l25gc,dpcm, cellclone, neutrino,mmlite,larrea2023corekube} aim to reduce control plane delays. L25GC~\cite{sigcomm22_l25gc} uses shared-memory communication, smart buffering at UPF and a quick failure recovery protocol. CoreKube~\cite{larrea2023corekube} leverages a stateless cellular core design for scalability and failure resilience. MMLite~\cite{mmlite} enables procedure prioritization but lacks an interface for edge apps to dynamically influence control plane processing during a latency-sensitive data session. The highlighted designs can be integrated into our \Name to augment performance, particularly in terms of failure recovery.


\section{Discussion}
\label{sec:discussion}

\noindent\textbf{Improving Mobility Hint:}
Figure~\ref{fig:eval:cdf_target_bs_prediction} shows \averageAccuracy correct predictions on average with a 100 ms mobility hint. Prediction accuracy can be further improved when apps (e.g., autonomous vehicles) inform the mobility handler about motion intentions based on destination settings, and the operator considers historical handover data and user GPS coordinates.

\noindent\textbf{Target BS Predictor Training:} 
We train the target BS predictor module offline with diverse mobility traces.
The offline model provides good performance
. However, telecom operators can also train it online on a BS or edge site to learn new mobility trends for more accurate predictions.

\noindent\textbf{Large State Sizes:} 
For apps with large state sizes (e.g., 100 MB), external data store read/write latency can become a bottleneck. These apps should cache large states locally, updating the external store only on mobility hints or handover initiation. In such scenarios, our proposed modifications may offer limited benefits.

\noindent\textbf{Always Sync Approach:}
Data stores, such as Redis~\cite{redis_store} and ZooKeeper~\cite{hunt2010zookeeper} allow state replication on the backup replicas for fault tolerance. However, always replicating the user state on the backup (i) can have a significant bandwidth overhead due to the typically high update rates of the latency-sensitive apps~\cite{vr-latency,zhang2017towards_online_game_updates_rate, ahmad_carmap, zhang2021emp}, and (ii) there may be less benefit of proactive replication due to the state soon becoming outdated due to frequent updates. 

\section{Concluding Remarks}
\label{sec:conclusion}
In this work, we addressed the problem of enabling stateful mobile edge applications. We identified the key architectural issues in enabling such demanding applications. We posit that there is a need to rethink coordination between edge apps and 5G networks. To this end, we proposed a new cross-layer system architecture, \Name. Our evaluations with real edge apps and simulations show that \Name takes a major step towards enabling such apps by significantly reducing app downtime during mobility. We hope our work will ignite an important discourse on enabling emerging mobile applications over 5G and beyond. 




\bibliographystyle{plain}
\bibliography{main,new}

\appendix
\clearpage
\section{Appendix}
\label{sec:appendeix_implementatioin}
In this section, we describe our setup to evaluate the existing application session migration tools.

\subsection{Application Session Migration Setup With Existing Tools}
\label{sec:existing_tools_detail}
Application handover for CarMap with existing tools, CRIU~\cite{criu} and LXC container, is performed between two Ubuntu 18.04 VMs (kernel version 5.0.0-36-generic) with 4GB RAM and 2 CPU cores. The link speed between the two VMs reported by the iperf3 is 1.62\,Gbps. 

We synchronize time on both machines using \textit{chrony}---an implementation of Network Time Protocol (NTP). We set up one machine as an NTP time server and configured the other machine to synchronize its time with the NTP server. The \textit{chronyc sources -v} reports a time skew of approximately +/- 150\,$\mu$s which is an acceptable threshold when measuring times in the order of hundreds of milliseconds.

\subsubsection{User-space tool CRIU}
\label{appendix:userspace_tool_criu}
We used CRIU v3.17.1 for a userspace application session migration.

\noindent\textbf{Reactive Migration:}
We use \textit{criu dump} command to checkpoint the application on the source machine, transfer the log files and checkpointed image to the destination machine using \textit{rsync} utility, and restore the application on the destination machine with \textit{criu restore} command. 

\noindent\textbf{Iterative Migration:}
We use \textit{criu pre-dump} command to take a snapshot of the running application. The \textit{criu pre-dump} command allows the application to run normally while taking a snapshot. We use 2 pre-dumps to get lower total migration and application downtime. We use \textit{criu dump} to checkpoint the memory changes since the last \textit{criu pre-dump} command was issued. We transfer the log files, intermediate snapshots, and checkpointed images to the destination machine using \textit{rsync} utility and restore it with \textit{criu restore} command.

\subsubsection{LXC Container}
\label{appendix:lxc_container}
We have used LXD version 3.0.3 for LXC container-based CarMap app migration. We use Ubuntu 14.04 as a base image due to the \textit{systemd} on Ubuntu 16.04 and the latest versions do not allow the CRIU to perform checkpoints.

\noindent\textbf{Reactive Migration:}
We use \textit{lxc move} command to perform reactive migration of an LXC container. 

\noindent\textbf{Iterative Migration:} We use the following container configuration for proactive application session migration~\cite{live_lxc_migration}:
\begin{itemize}
    \item  Proactive container migration is performed by setting \textit{migration.incremental.memory = true}.
    \item The percentage of memory that needs to be synchronized before the final transfer is set to 80 with the command\\ \textit{migration.incremental.memory.goal = 80}.
    \item The maximum number of iterations or pre-dumps being set to 20 with command \\\textit{migration.incremental.memory.iterations = 20}.
\end{itemize}

The average total migration time for both reactive and proactive migration is $\approx$27 seconds.

\section{Appendix}
\label{sec:appendeix_additional_evl_results}
In this section, we provide some additional evaluation results.

\subsection{Inference Time for Target BS Prediction Pipeline}
\label{sec:appendix_inference_time}


\begin{figure}[]
\centering
\includegraphics[width=\columnwidth]{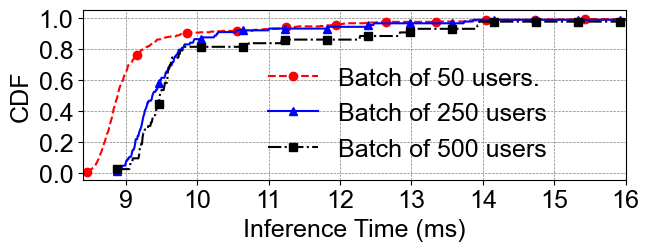}
\vspace{-0.2in}
\caption{Inference time of the target BS prediction pipeline when varying the batch sizes.}
\label{fig:bs_inference_time}
\end{figure}

\begin{figure}[]
\centering
\includegraphics[width=\columnwidth]{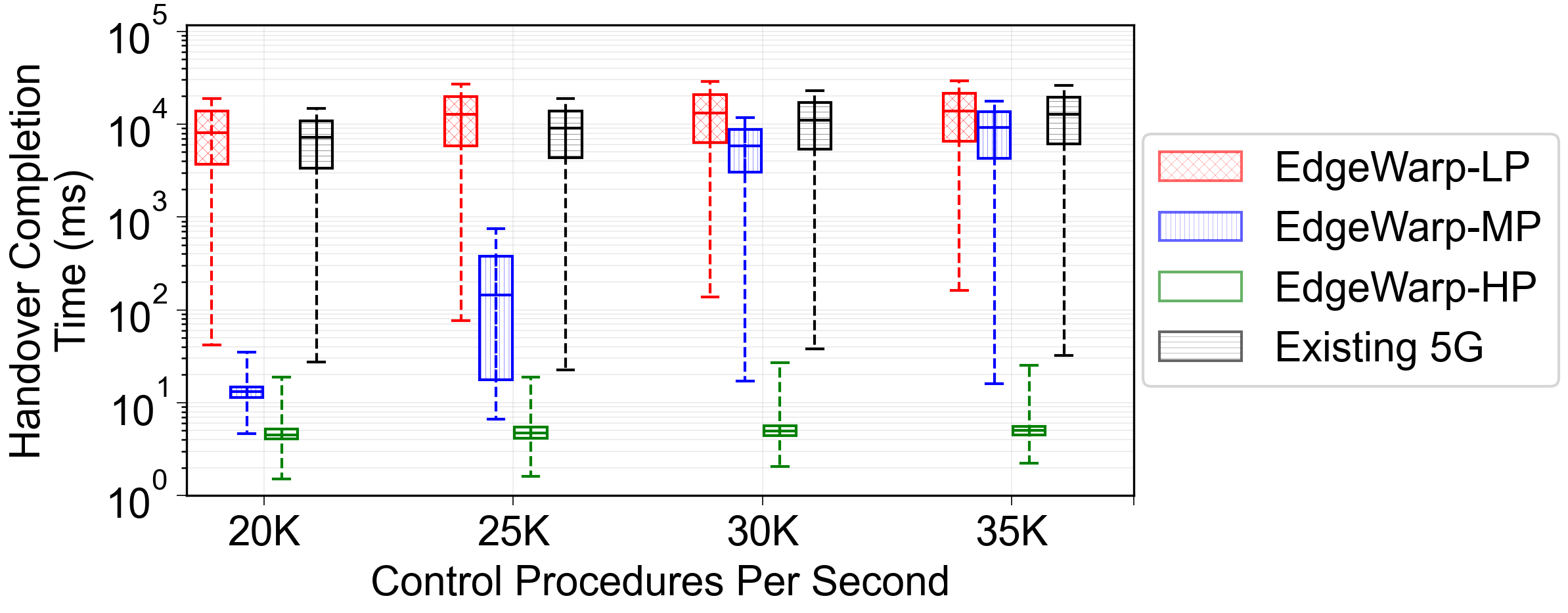}
\caption{A comparison of \handover completion time for \Name and \Old  with uniform control traffic. Traffic split 60\% \lowPriorityIot, 28\% \mediumPriorityIot, and 12\% for \highPriorityIot, as projected in \cite{iot_traffic_distribution}. }
\label{fig:app_impact_on_uniform_ho_pct}
\end{figure}

Figure~\ref{fig:bs_inference_time} shows the CDF of the inference time for the target BS prediction pipeline. We calculate the inference time on Google Colab with NVidia Tesla K80 GPU. We compute inference time when the target BS prediction is performed for a batch of 50, 250, and 500 users. The figure shows that median inference time is 8.7\,ms and 9.5\,ms when target BS prediction is performed for a batch of 50 and 500 users, respectively.

\subsection{App-Aware Control Plane Impact on Handover Completion Time}
\label{sec:appendix_control_plan_bursty_traffic}

In this section, we show the impact of the App-aware control plane on \handover completion time when the cellular control plane is handling a varying number of control procedures--a set of related control plane updates~\cite{neutrino_ton, dpcm} such as \handover.

Figure \ref{fig:app_impact_on_uniform_ho_pct} shows \handover completion time when the 5G control plane is serving a varying number of control procedures per second. The figure shows that the median \handover completion time for high-priority latency-sensitive applications (\highPriorityIot) remains around 4.5\,ms even when the load on the control plane increases from 20K to 35K control procedures per second. This is two orders of magnitude less than the case when the users are served by \Old. As a trade-off, we observe around 
 \uniformLowOverDefaultMedium increase in \handover completion time for low priority delay tolerant applications (\lowPriorityIot) as compared to the  \defaultPriority. However, this is an acceptable trade-off as the \lowPriorityIot can tolerate a delay of up to 10\,s~\cite{latency_nbiot_catm, iot_in_5g_era_cisco, 3gpp_evaluation_mmtc}
and it is still < 1\,s. In all load scenarios, we observe the medium priority delay tolerant applications (\mediumPriorityIot) has a lower \handover completion time as compared to the \defaultPriority which suggests no degradation in the performance of \mediumPriorityIot.

\begin{figure}[]
\centering
\includegraphics[width=\columnwidth]{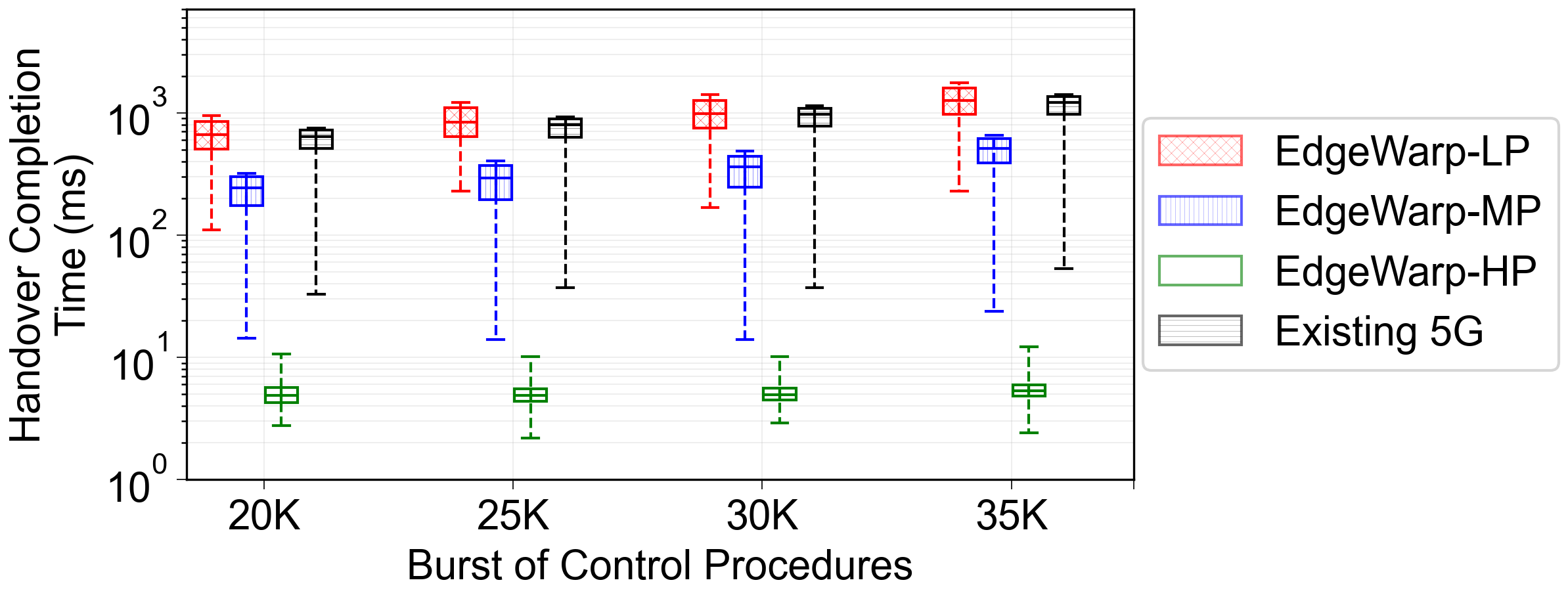}
\vspace{-0.25in}
\caption{A comparison of \Name's \handover completion time and \Old with bursty control traffic. Traffic split 60\% \lowPriorityIot, 28\% \mediumPriorityIot, and 12\% for \highPriorityIot, as projected in \cite{iot_traffic_distribution}.}
\vspace{-0.15in}
\label{fig:app_impact_on_bursty_ho_pct}
\end{figure}

To test at scale, figure \ref{fig:app_impact_on_bursty_ho_pct} shows \handover completion time when the 5G control plane is serving bursty IoT control traffic. The figure shows that irrespective of the traffic load, medium time for \handover completion  remains around 5\,ms for high-priority latency-sensitive applications (\highPriorityIot). This is more than two orders of magnitude lower than the case the users are served by \Old. The Figure shows up to \burstyLowOverDefaultMedium increase in \handover completion time for low priority delay tolerant applications (\lowPriorityIot) as compared to  \defaultPriority which is an acceptable trade-off as \lowPriorityIot can tolerate a delay of up to 10\,s~\cite{latency_nbiot_catm, iot_in_5g_era_cisco, 3gpp_evaluation_mmtc} which is round 1\,s in this case. The figure also shows that in all load scenarios, the performance of medium priority delay tolerant applications (\mediumPriorityIot) is better than the \defaultPriority.


\begin{figure}[]
\centering
\includegraphics[width=\columnwidth]{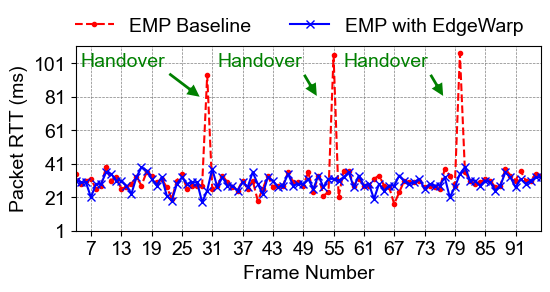}
\vspace{-0.3in}
\caption{Improvement in message RTT during an app handover when using EMP with \Name as compared to EMP Baseline. Setting row 1 of table~\ref{table:eval:emp_state_migration_time}.}
\vspace{-0.15in}
\label{fig:eval_emp_rtt}
\end{figure}

\subsection{Round Trip Time Latency on Vehicle for EMP}
\label{appendix_rtt_emp}
Figure \ref{fig:eval_emp_rtt} shows message RTT measured on the vehicle when three app handovers are performed during an experiment. The figure shows more than 3$\times$ improvement in EMP with \Name as compared to the baseline. These improvements are attributed to the mobility hint-based proactive application state synchronization provided by the \Name's modified \redis data store.

\section{Appendix}
\label{sec:appendeix_target_bs_prediction}
In this section, we provide some additional details about the target BS prediction pipeline of Figure~\ref{fig:target_bs_prediction_architecture}.

\subsection{Data Sets Description}
\label{appendix_dataset_description}

We use the following data sets to train and evaluate our target BS prediction pipeline:

\begin{enumerate}
    \item High-speed trains data set: This data set~\cite{hst_sigcomm_dataset} was collected in high-speed trains (for a study published in ACM SIGCOMM~\cite{li2020beyond}) where faster mobility can be a challenge for emerging edge applications. This data set has a total of \hsrHandovers handover events.

    \item Driving data set: The data set~\cite{incp_dataset} was collected during driving tests for a study published in IEEE INCP~\cite{incp_li2021reconfiguring}. This data set has a total of \incpHandovers handover events.\footnote{This data set is not used for training the LSTM network to avoid over-fitting.}

    \item Miscellaneous data set: This data set includes multiple traces published to MobileInsight~\cite{mobile_insight_logs} website with no explicit mention of mobility scenarios. This data set has a total of \miscHandovers handover events.
\end{enumerate}

We used 10\% of the above data for training, 3\% for validation, and 87\% for testing.

\subsection{Feature Vector for Target BS Prediction}
\label{appendix_feature_vector}

We pre-process the data sets described in section \ref{appendix_dataset_description} (having \totalHandovers handover events) to convert them to feature vectors, as shown in figure~\ref{fig:feature_Vector}. The data sets are sampled on $\approx$50\,ms interval. A feature vector includes the last \textit{N} (configurable) samples of Reference Signal Received Power (RSRP) for target BS, Reference Signal Received Quality (RSRQ) for target BS, RSRP 
of the target BS relative to source BS, and RSRQ of the target BS relative to source BS. 

Figure~\ref{fig:feature_Vector} also highlights the timeline of some important events generated in our target BS prediction pipeline in figure~\ref{fig:target_bs_prediction_architecture}. The figure shows that the target BS predictor module predicts handover at sample number 3. At sample number 5, \Name can correctly predict the target BS for handover. The actual 5G handover decision by the source BS is taken at sample number 8. Here \Name can predict the target BS two samples ($\approx$100\,ms) before an actual handover.

\begin{figure}[]
\centering
\includegraphics[width=\columnwidth]{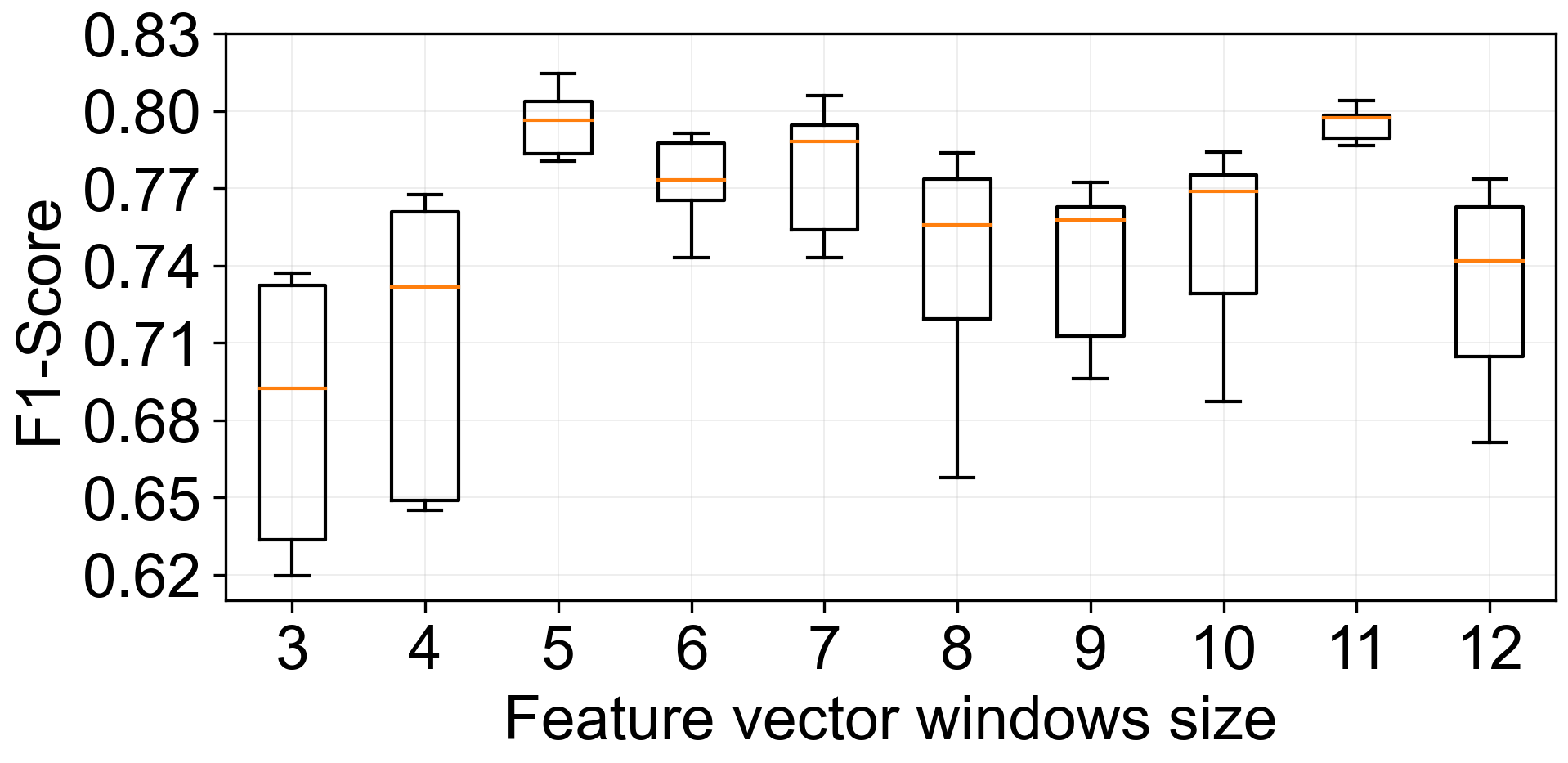}
\vspace{-0.25in}
\caption{Feature vector windows size vs. F1 score.}
\label{fig:feature_vector_windows_size}
\end{figure}

Figure~\ref{fig:feature_vector_windows_size} shows F1-scores of 10 models trained on feature vectors with different window sizes. The figure shows Windows size 5 performs the best in terms of the early target BS prediction and F1-score. All our evaluations in the paper are performed with a window size of 5.



\begin{figure}[]
\centering
\includegraphics[width=\columnwidth]{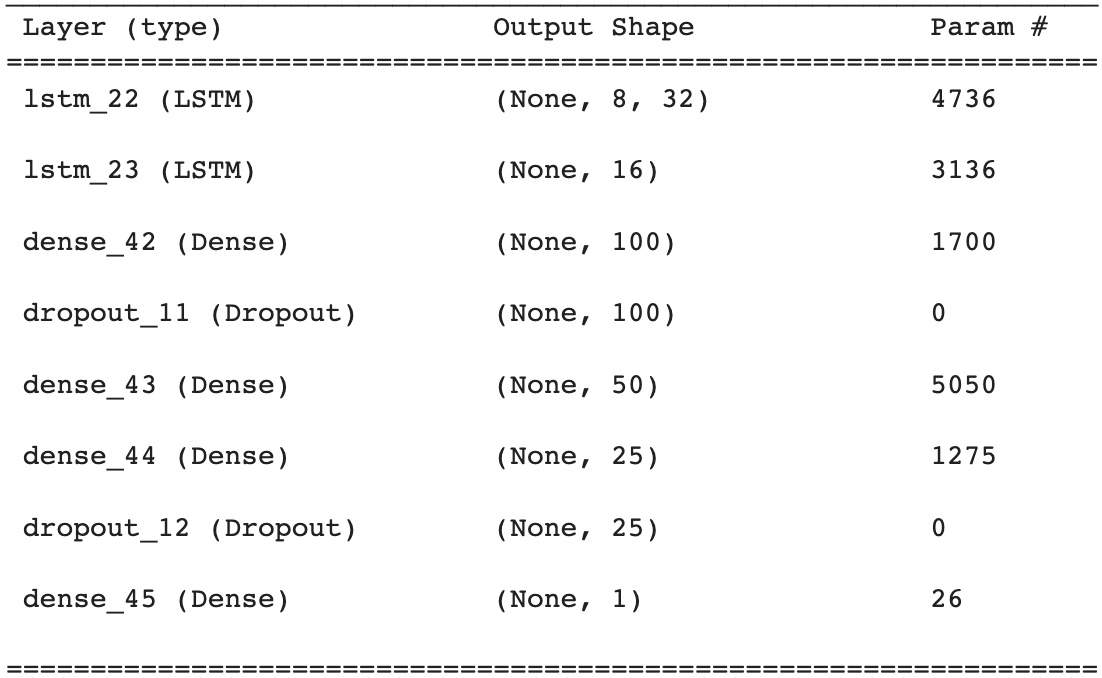}
\vspace{-0.2in}
\caption{A summary of the neural network for target BS prediction module of figure~\ref{fig:system_architecture_diagram}.}

\label{fig:bs_prediction_neural_network}
\end{figure}

\subsection{LSTM Binary Classifier}
\label{appendix_lstm_binary_classifier}

Figure~\ref{fig:bs_prediction_neural_network} shows a summary of the LSTM network used for target BS predictions. The figure shows the LSTM network has a total of 8 layers. The figure also shows trainable parameters in every layer of the LSTM network.

\begin{figure*}[b]
\centering
\includegraphics[width=\textwidth]{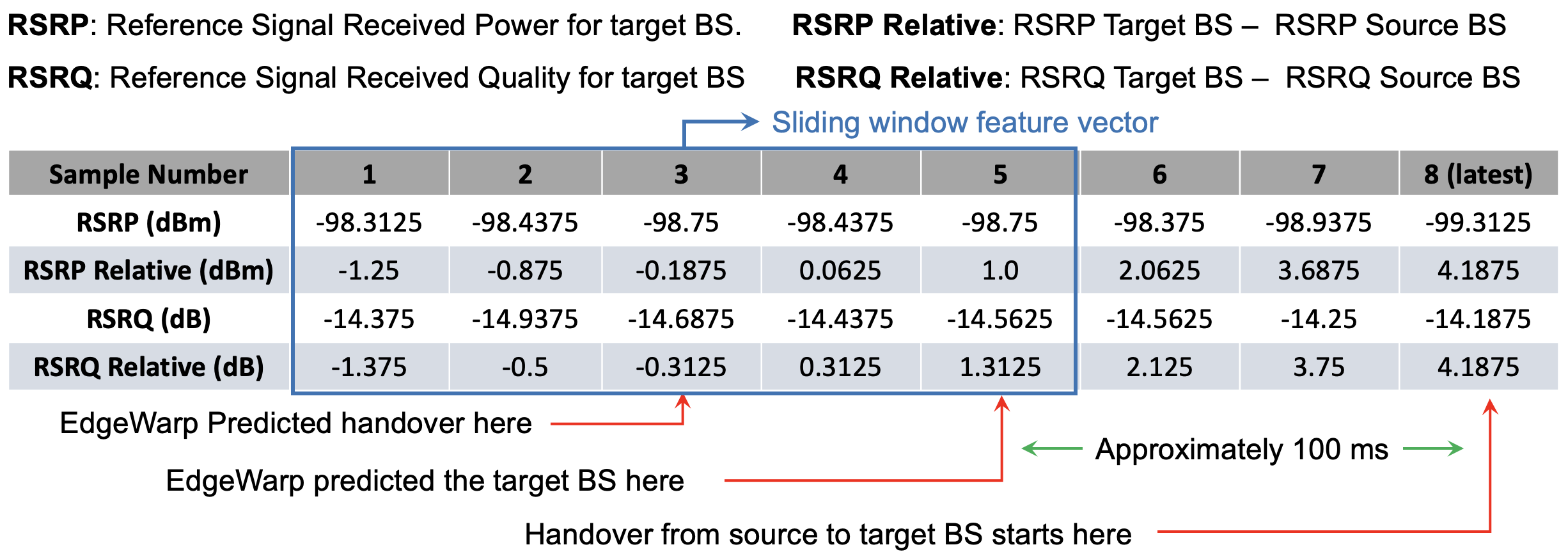}
\vspace{-0.2in}
\caption{A snapshot of the sliding windows (size 5) feature vector for source and one of the potential target BSs. The samples shown in  figure~\ref{fig:system_architecture_diagram}  are sampled on $\approx$50\,ms interval.}
\label{fig:feature_Vector}
\end{figure*}

\section{Appendix}
\label{appendix_d}

In this section, we show the sequence of steps an app-aware 5G control plane performs to provide priority control plane processing for a UE when running a latency-sensitive application session.

\noindent\textbf{{Sequence of Operations:}}
Figure \ref{fig:system_architecture_diagram} shows a MEC system hosting multiple latency-sensitive edge apps. We show the steps of how an application can influence control plane operations for a UE.
\begin{enumerate}[leftmargin=*]
    \item When a UE transitions from idle to connected mode~\cite{3gpp_nas}, the AMF provides the required session state~\cite{3gpp_architecture} to the BS. The AMF informs the BS whether a UE is a delay-tolerant low-end device (\lowPriorityIot) or a high-end device (\mediumPriorityIot or \highPriorityIot).
    
    \item The UE starts a data session with a latency-sensitive edge app hosted at the MEC platform. The edge app acts as an AF.

    \item The edge app sets the highest priority for the UE with a call to \textit{Set\_Priority} API (in table \ref{tab:api_for_app_influence_on_cpf}) on AMF with the UE IP address as the UE's unique ID. The AMF (having access to the complete UE's control plane state) translates the UE's IP address to the user unique IDs \textit{RAN UE NGAP ID} and \textit{AMF UE NGAP ID} at the BS and AMF, respectively. The AMF sets the highest priority for the UE on RAN by calling the \textit{Set\_Priority} API (table~\ref{tab:api_for_app_influence_on_cpf}) exposed by RAN with \textit{RAN UE NGAP ID} as a unique UE's ID.
    
    \item With a call to the above APIs, the BS and AMF add the UE to the high-priority list maintained for \highPriorityIot.
    
    \item When the UE finishes the data session with the latency-sensitive edge app, the edge App reverts its priority to default with a call to \textit{Revert\_Priority} API (in Table \ref{tab:api_for_app_influence_on_cpf}) exposed by AMF. Consequently, the AMF calls \textit{Revert\_Priority} on RAN to revert the UE's priority to default.
\end{enumerate}

\end{document}